\documentstyle[11pt,aaspp4]{article}
\newcommand{\Rmin}{R_{\rm min}}
\newcommand{\Rmax}{R_{\rm max}}
\newcommand{\mean}[1]{\left\langle#1\right\rangle}
\newcommand{\lop}{L_{{\rm op}}}
\newcommand{\eg}{e.$.\!$g.\ } 
\newcommand{\lr}{L_{\rm r}}
\newcommand{\be}{\begin{equation}}
\newcommand{\ee}{\end{equation}}
\newcommand{\etal}{{\it et al.\ } }
\newcommand{\ie}{i$.\!$e$.\!$, }
\newcommand{\gne}[2]{\ \vphantom{S}^{\raise-0.5pt\hbox{$\scriptstyle#1$}}_{\raise0.5pt \hbox{$\scriptstyle#2$}}}
\newcommand{\gneq}{\gne > \sim\,}
\newcommand{\lneq}{\gne < \sim\,}
\newcommand{\onlyten}[1]{10^{#1}}
\newcommand{\percent}{\unit{percent }}
\newcommand{\unit}[1]{\,{\rm #1}}

\newcommand{\ghz}{\unit{GHz}}

\newcommand{\whzs}{\unit{W\,Hz^{-1}\,str^{-1}}}
\newcommand{\mjy}{\unit{mJy}}

\newcommand{\ergsh}{\unit{erg\,sec^{-1}\,Hz^{-1}}}
\newcommand{\aop}{\alpha_{\rm op}}
\newcommand{\ar}{\alpha_{\rm r}}
\newcommand{\fr}{F_{\rm r}}
\newcommand{\fop}{F_{\rm op}}
\newcommand{\philopz}{\Phi_{\rm op}(\lop,z)}

\newcommand{\philr}{\Phi_{\rm r}(\lr)}
\newcommand{\phir}{\Phi_{{\rm R}}(R)}
\newcommand{\philoplrz}{\Phi(\lop,\lr,z)}
\newcommand{\dlop}{dL_{\rm op}}
\newcommand{\dlr}{dL_{\rm r}}
\newcommand{\dvz}{dv(z)}
\newcommand{\hubble}[1]{H_0={#1}\unit{km\,sec^{-1}\,Mpc^{-1}}}

\newcommand{\qz}{q_0}
\newcommand{\labequn}[1]{\label{eq:#1}}
\newcommand{\labfig}[1]{\label{fig:#1}}
\newcommand{\labsecn}[1]{\label{sec:#1}}
\newcommand{\labsubsecn}[1]{\label{subsecn:#1}}
\newcommand{\equn}[1]{Equation~\ref{eq:#1}}

\newcommand{\fig}[1]{Figure~\ref{fig:#1}}

\newcommand{\subsecn}[1]{subsection~\ref{subsecn:#1}}

\begin{document}
\title{A Study of Quasar Radio Emission from the VLA FIRST Survey}
\author{Yogesh Wadadekar,\altaffilmark{1} and Ajit Kembhavi \altaffilmark{2}}
\affil{Inter University Centre for Astronomy and Astrophysics, Post Bag
4, Ganeshkhind, Pune 411007, India}
\altaffiltext{1}{yogesh@iucaa.ernet.in}
\altaffiltext{2}{akk@iucaa.ernet.in}
\begin{abstract}
Using the most recent (1998) version of the VLA FIRST survey radio
catalog, we have searched for radio emission from 1704 quasars taken
from the most recent (1993) version of the Hewitt and Burbidge quasar
catalog.  These quasars lie in the $\sim$5000 square degrees of sky
already covered by the VLA FIRST survey. Our work has resulted in
positive detection of radio emission from 389 quasars of which 69
quasars have been detected for the first time at radio wavelengths.
We find no evidence of
correlation between optical and radio luminosities for optically
selected quasars. We find indications of a bimodal distribution of
radio luminosity, even at a low flux limit of $1 \mjy$. We show that
radio luminosity is a good discriminant between radio loud and radio
quiet quasar populations, and that it may be inappropriate to make
such a division on the basis of the radio to optical luminosity
ratio. We discuss the dependence of the radio loud fraction on optical 
luminosity and redshift.

\end{abstract}
\section{INTRODUCTION}

It has been well known for some time that only about 10\% of quasars
are radio loud, with radio luminosity comparable to optical
luminosity. This is surprising, because over a very wide wavelength
range from ~100 $\mu$m through  X-ray wavelengths, the properties of
radio loud and radio quiet quasars are very similar. The presence or
absence of a radio component may be a pointer to different physical 
processes occurring in the two types of quasar, but
it is not yet clear as to what these processes are.

The relationship between quasar radio and optical emission was
initially studied using radio selected objects, which generally had
high radio luminosities because the early radio surveys had relatively
high limiting radio fluxes. Sandage (1965) showed that not all quasars
are powerful radio emitters, and that a substantial population of {\em
radio quiet} quasars exists, undetectable at high radio flux
levels. Since then, in addition to radio surveys, radio follow up
observations of large surveys conducted in the optical have been used
to study the radio properties of quasars (eg. Sramek \& Weedman 1980;
Condon \etal 1981; Marshall 1987; Kellerman \etal 1989; Miller, Peacock \& Mead 1990). Such
targeted radio observations, of quasars selected by other means,
typically go deeper than the large radio surveys, as a result of which
the median radio luminosity of these samples is lower. Taken together,
these two survey methods have detected quasars with a range of more
than 6 orders of magnitude in radio luminosity, but the populations
detected by the two methods come from different regions of the overall
radio luminosity distribution.

The radio emission from quasars can be used to divide them into two
classes: a radio loud population where the ratio $R$ of radio to
optical emission is greater than some limiting value $R_{\rm lim}$ and
a radio quiet population with $R < R_{\rm lim}$.  Such a separation is
commonly employed in the literature dealing with the radio properties
of quasars, with $R_{\rm lim}=1$ or $R_{\rm lim}=10$ (eg. Kellerman
\etal 1989; Visnovsky \etal 1992; Stocke \etal 1992; Kellerman \etal
1994). Alternately, the separation between radio loud and radio quiet
quasars, may be defined by their radio luminosity.  Such a criterion
has been advocated by Miller, Peacock \& Mead (1990), who noticed that
for a sample of optically selected quasars, which spanned a wide range
of optical luminosity but a narrow range of redshift, there was no
correlation between their optical and radio luminosity.  This implied
that the distribution of $R$ was optical luminosity dependent, thus making it
unsuitable as the discriminant between radio loud and radio quiet
populations.  Miller \etal found that the distribution of radio
luminosity was highly bimodal, and from an examination of the
luminosities of radio detections and upper limits accepted a $5\ghz$
limiting radio luminosity of $\onlyten{25}\whzs$ (we use
$\hubble{50}$, $\qz=0.5$, quasar radio spectral index $\ar=0.5$ and
optical spectral index $\aop=0.5$ throughout this paper) as the
dividing line between radio loud and radio quiet quasars.
 
The gap in the radio luminosity function of the two populations is
pronounced, with very few objects occupying the
region between quasars that are radio loud and those that are radio
quiet. The detection technique used to find quasars from these two
populations are also different. An overwhelming majority of radio loud
quasars have been first detected in the radio and then confirmed using
optical spectroscopy, while radio quiet quasars have been detected
using optical, X-ray or other techniques. An important question in
such a situation is: are radio quiet and radio loud quasars indeed two
physically different populations, or is the distinction merely an
artifact caused by selection biases in the detection techniques?
Previous efforts at answering this question have been plagued by the
small size of the datasets and their incompleteness. Most radio
observations of optically selected quasars have lacked the sensitivity
to detect their radio emission. There have been a few high sensitivity
radio surveys (eg. Hooper \etal 1996, Kukula \etal 1998) but the size
of their samples is quite small.  The VLA {\em Faint Images of the
Radio Sky at Twenty centimeters} (FIRST) survey (Becker \etal 1995;
for more upto date information see the FIRST survey homepage at {\em
http://sundog.stsci.edu/}) allows us to address this question
meaningfully, by combining a large sky coverage with a low flux limit
of $1 \mjy$ at 20 cm. This ongoing survey, when completed will cover
10,000 square degrees around the North Galactic Cap, the same area of the sky to be
surveyed by the Sloane Digital Sky Survey (SDSS;
http://www.sdss.org/). To date, data for approximately one half of the
eventual sky coverage have been released.

FIRST allows us to address the issue of quasar bimodal radio
luminosity distribution in two different but complementary
ways. Firstly, optical identifications of FIRST sources using large
optical surveys such as the Palomar Observatory Sky Survey (POSS)
provide a large database of quasar candidates, whose true nature can
then be verified spectroscopically. Several such efforts (eg. Gregg et
al 1996; Becker \etal 1997) are currently underway. Secondly, the
large area covered by the FIRST survey allows us to look for radio
emission from a significant fraction of already known quasars and
correlate their radio properties with other observables. In the
present paper, we have used this approach to determine the radio
properties of quasars from the catalog of Hewitt \& Burbidge (1993,
hereafter HB93).

Such an approach has also been taken, though with a different radio
survey and quasar catalog, by Bischof and Becker (1997, hereafter
BB97) who compared positions of radio sources from the NVSS radio
survey (Condon \etal 1998), with the positions of 4079 quasars from
the Veron catalog (Veron-Cetty and Veron 1991).  They detected radio
emission from 799 quasars, of which 168 were new radio detections.

The FIRST survey has better sensitivity and resolution than the NVSS,
but covers a smaller area. There is a small area of overlap between
NVSS and FIRST.  The FIRST survey, which is being carried out with the
VLA in its B-configuration, has excellent astrometric accuracy of
$\sim1\arcsec$ (90\% error circle) and a 5 sigma sensitivity of
$\sim1\mjy$. This compares favorably with the D-array NVSS, which has
a beam size of 45 arcsec and a 5 sigma sensitivity of $\sim2.4\mjy$.
FIRST has a smaller beam size than NVSS, and so it is expected to have
better sensitivity to point sources.  We look for radio emission from
the 1704 quasars from HB93 ($\sim23\%$ of the quasars listed therein)
which lie in the area covered by the FIRST survey.  This set of
quasars is not statistically complete in any sense. Wherever
appropriate, we distinguish between radio selected quasars and those
selected by other means.

\section{RADIO/OPTICAL COMPARISONS}

We compare the positions of quasars in HB93 to the positions of radio
sources in the FIRST radio source catalog (February 4, 1998 version
available at {\em http://sundog.stsci.edu/}), and calculate the
angular separation between each quasar and each FIRST source. About
4\% of sources in the FIRST catalog have been tagged as possible
sidelobes of bright sources. Of these, $<$10\% are real sources and
considerably less than 1\% of the unflagged sources in the catalog are
sidelobes (White \etal 1997). We have excluded these flagged sidelobe
sources from our cross correlation. We are then left with a total of
421,447 unflagged sources in the northern and southern strips,
covering a total area of about 4760 square degrees. Of these, 368,853
sources lie in the northern strip while 52,594 sources are in the
southern strip. On an average, there are 88.54 FIRST sources per
square degree of sky.

Our quasar sample consists of 1704 quasars from HB93, that lie in the
area covered by FIRST. We have excluded the BL Lacs listed in the
catalog from the present work. In HB93, the authors use a simple
selection criterion for quasars. Any object that is starlike (with or
without fuzz) and has redshift $z \ge 0.1$ is called a quasar; and is
included in the catalog. The positions listed in the catalog are for
the optical object, most of which are taken from the identification
paper or from the paper containing the redshift measurement. If a quasar
is very close to a bright galaxy, and the quasar coordinates are not
available in the literature, the galaxy coordinates have been listed
by HB93 for the quasar position.

\begin{figure}
\epsscale{0.6}
\plotone{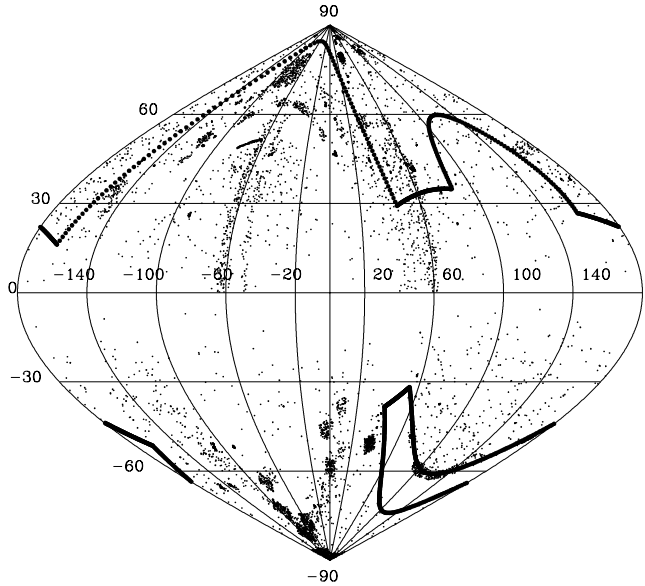}
\caption{The approximate boundaries, in galactic coordinates, of the areas covered by the FIRST
survey are indicated by bold points. The
northern and southern strips are separately shown. The dots indicate
quasars from the Hewitt and Burbidge (1993) catalog.}
\labfig{quasky}
\end{figure}

In \fig{quasky} we show the distribution of the HB93 quasars on the sky in galactic coordinates. The apparent clustering of
known quasars is due to the limited solid angle covered by deep quasar
surveys (most of them optical) that have uncovered the largest number
of quasars. The approximate area for which FIRST survey data has been
released is also marked in the figure. FIRST has currently covered an
area of 4150 square degrees around the North Galactic Cap in addition
to two narrow strips totaling about 610 square degrees near the South
Galactic Cap. The southern strip has a peculiar shape and the box
shown here is a very approximate representation. Detailed sky coverage
maps are available at the FIRST homepage.

In order to find coincidences between HB93 and FIRST sources, we begin
with a search circle of radius $300\arcsec$ centered on each HB93 quasar,
and look for FIRST radio sources within this circle.  When there is
more than one FIRST source in the search circle, we tentatively accept
all such sources as matches.  In \fig{sephist} we show a histogram of
the angular separation between the HB93 quasars and the FIRST sources
found in the search circles.

\begin{figure}
\plotone{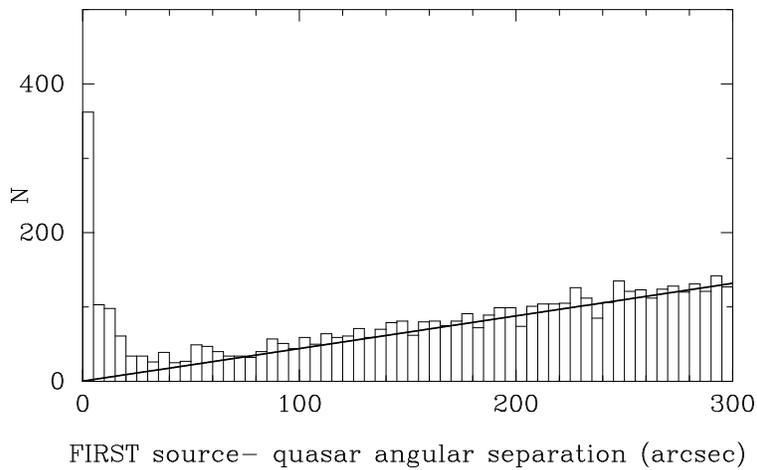}
\caption{A histogram of the angular separation between the Hewitt and
Burbidge (1993) quasars and the corresponding FIRST source. The straight line
is the number of chance matches expected, for a search area of radius
shown on the X-axis,  if the FIRST sources were
randomly distributed in the sky.}
\labfig{sephist}
\end{figure}

The angular auto correlation function for FIRST has shown that 35\% of
sources have resolved structure on scales from 2--30 arcsec (Cress et
al. 1996). Since our aim in this work is to only look for radio
emission from the compact (flat spectrum) component of quasars, we
have considered only quasars which contain at least one FIRST source
within 10 arcsec of them. This would make us miss out on some quasars
which may have elaborate extended radio structure, but a core emission
lower than the FIRST flux limit.  To see which of the radio sources
found can be accepted as true identifications, we estimate the
quasar--FIRST source chance coincidence rate for a random distribution
of FIRST survey sources. For a random distribution, the chance
coincidence rate is directly proportional to the area of sky covered
by the search circle around each quasar, \ie the the square of the
search radius. The straight line in \fig{sephist} is the expected
number of chance coincidences between quasars and FIRST sources, in
annuli of radius shown on the abscissa and width 5 arcsec around the
quasars.  It is seen from the figure that the expected number of
chance coincidences closely matches the number of actual coincidences beyond
about 40 arcsec, indicating that most FIRST sources found more than 40
arcsec away from a quasar are chance coincidences. On the other hand,
matches within 10 arcsec are mostly real (less than 1\% chance
identification probability). 

We therefore choose a search circle of
radius 10 arcsec and count all matches found within this radius as true
matches. All subsequent discussion about the radio properties of
quasars only uses matches obtained with this search radius.

The positions of quasars listed in HB93 have astrometric errors of a few
arcseconds or more in some cases.  In such cases,
there will be  missed matches when the positional error places a HB93 quasar
outside the 10 arcsec search radius around the FIRST radio source with which
it is actually associated.  Some  of these missed  can be recovered
by using the accurate positions of star like objects from the
USNO-A2.0 catalog (Monet \etal 1996), which is an all sky
astrometric and photometric catalog of over 500 million starlike objects.
For this purpose we considered FIRST radio sources which had a HB93 quasar
in an annulus of inner radius 10 arcsec and outer radius 20 arcsec around it.
We cross correlated the positions of such radio sources with starlike sources 
from the USNO-A2.0 catalog.  The search radius used for this purpose was of
3 arcsec, whish is three times  the RMS uncertainty in the first
survey positions (the positions in USNO-A2.0 are known to better than this accuracy).  When an
USNO-A2.0 object is found in this circle we compare its blue magnitude with
the blue magnitude of the corresponding HB93 quasar.  When the  difference
$\delta m$ was less than one magnitude, we considered the USNO-A2.0 object
and the HB93 quasar to be the same object.   

We have a total of 158 FIRST sources with a HB93 quasar within the
10-20 arcsec annulus around it.  Out of these 158 sources, 16 had a
USNO-A2.0 source within a 3 arcsec circle around it, and of these 8 had
blue magnitudes which passed our criterion.  We accepted the 
corresponding 8 HB93 quasars as valid matches with FIRST survey sources,
and added these to our list of 381 radio detections mentioned above. 
 
We find that positions of radio selected quasars match the FIRST
source positions better than the non-radio selected quasars. This is
because the radio positions have been more accurately
determined than optical positions. Accurate astrometry on optically
selected quasars is often not available and quasar positions are
computed approximately using finding charts. The mean astrometric
error, in non-radio selected quasar surveys, is typically a few
arcseconds. In high resolution radio selected surveys, the astrometric
error is often less than an arcsecond.  About 12\% of quasars detected
have more than one (usually two) FIRST sources within the search
circle of 10 arcsec radius. In such cases, we have used {\em all} the
FIRST sources associated with the quasar in our analysis. This is
because generally the combined error in quasar and FIRST source positions is too large to
allow us to reliably determine which of the two radio sources actually
corresponds to the quasar core.  There are $\sim$1320 non detections,
amongst the HB93 quasars covered by FIRST, and we assign an upper
limit of $1 \mjy$ to their radio flux at 1.4 \ghz.
 
Table 1 provides a summary of our radio detections.  
The radio and optical properties of the quasars with FIRST
detections (which includes all the new detections in the radio) are
summarized in Table 2. Detections in the radio reported after the
quasar catalog was published (mostly in BB97) are mentioned in the
last column. Those quasars which do not have the letter R in the
selection technique code and do not have a recent radio detection
mentioned in the last column may be considered to be the new detections. There are 69 such
quasars in Table 2. The last eight entries are the additional list of matches
obtained using a correlation with the USNO-A2.0 catalog. These 8
matches were obtained using an indirect comparison technique, and have 
 {\em not} been  used in the statistical correlations reported
in subsequent sections.

\subsection{1343+266: not a gravitationally lensed quasar?}

This is a close pair of quasars with identical redshift, similar
spectra and separated by only $\sim$10 arcsec. Detailed spectroscopic
observations have shown qualitative (eg. presence of certain lines) as
well as quantitative (eg. ratio of line strengths) {\em differences}
between the two quasars, strengthening the claim that this is {\em
not} a gravitationally lensed pair, but a physically associated pair
of quasars, possibly residing in a cluster of galaxies at $z=2.03$
(Crampton \etal 1988; Crotts \etal 1994). The optical luminosities
are comparable, with 1343+266B having a luminosity higher by about 5\% 
than 1343+266A.  We find radio emission from only one of the
quasars: 1343+266B has a flux of $8.9 \mjy$. The separation between the
gravitationally lensed quasar and the FIRST source is 2.18 arcsecond,
which is consistent with an error of $\sim1\arcsec$ each in the quasar
optical position and the FIRST radio position.  There is no radio
emission associated with 1343+266A at the FIRST flux limit of $1
\mjy$, because the FIRST source associated with 1343+266B is 7.4
arcsec away, too far to be associated with 1343+266A, considering the
extremely accurate astrometry done for this well studied pair of
quasars. There is no other FIRST source associated with 1343+266A.
This implies that the radio luminosity of 1343+266B is at least 8.9
times higher than that of 1343+266A, in sharp contrast to only 5\%
difference in their optical luminosity. Such radio detection is strong
evidence that the pair is {\em not} gravitationally lensed.

\section{RADIO AND OPTICAL PROPERTIES}
\labsecn{radopprop}

\subsection{Bivariate luminosity functions}
\labsubsecn{bivariate}

The number of quasars with optical luminosity in the range ($\lop,
\lop+\dlop$), radio luminosity in the range ($\lr, \lr+\dlr$) and
redshift in the range ($z, z+dz$) is in general given by $\philoplrz
\dlop\dlr\dvz$, where the {\it luminosity function} $\philoplrz$ is
the comoving number density of quasars for unit ranges of the
respective luminosities, and $\dvz$ is a comoving volume element at
$z$. If the radio and optical luminosities are independently
distributed, it is possible to separate the luminosity function as

\be
\philoplrz \dlop \dlr \dvz = \philopz\dlop\dvz\philr\dlr.
\labequn{lumfn1}
\ee

In this case there will be no correlation between the optical and
radio luminosities of the quasars described by \equn{lumfn1}.

Another form of the bivariate luminosity 
function extensively considered in the literature has been
\be 
\philoplrz\dlop\dlr\dvz = \philopz\dlop\dvz\phir dR,
\labequn{lumfn2}
\ee
where 
\be
R = \frac{\lr}{\lop} = \frac{\fr}{\fop}(1+z)^{\ar-\aop}
\labequn{lumfn3}
\ee
and $\fop$ and $\fr$ are the optical and radio flux densities at some
fiducial points in the spectrum, which we will take to be at $2500{\rm\AA}$
and $5\ghz$ respectively.  Since we take $\aop=\ar=0.5$, the ratio of
the luminosities is simply equal to the ratio of the fluxes.  It is
assumed here that the distribution of $R$ is independent of the other
variables.  This form of the luminosity function was first introduced
by Schmidt (1970) to describe the bivariate luminosity distribution of
3CR radio quasars. For a given optical luminosity $\lop$, it follows
from \equn{lumfn3} that the radio luminosity ranges from $\Rmin\lop$
to $\Rmax\lop$ for $R$ in the range $\Rmin<R<\Rmax$ and the mean radio
luminosity is given by
 
\be
\mean{\lr} = \mean{R}\lop, ~~~ \mean{R}=\int_{\Rmin}^{\Rmax}R\phir\,dR,
\labequn{lumfn4}
\ee
with the function $\phir$ being normalized to unity.  The luminosity
function in \equn{lumfn2} therefore implies that the mean radio
luminosity increases with the optical luminosity.

In the following sections we will see whether the data from the FIRST
survey is consistent with either of the two forms of luminosity
function.  In our discussion, we will use the following nomenclature
to refer to different classes of quasars:
\begin{itemize}
\item RSQ: Radio selected quasars, 
\item OSQD: non-radio (mostly optical) selected quasars detected by FIRST, 
\item OSQU: optically selected quasars with radio upper limits. 
\end{itemize}
The OSQD and OSQU include a few X-ray selected quasars, but their
numbers are too small to warrant separate treatment. All radio
selected quasars lying in the area covered by FIRST have radio
emission higher than the FIRST limit of 1 mJy.

\begin{figure}
\plotone{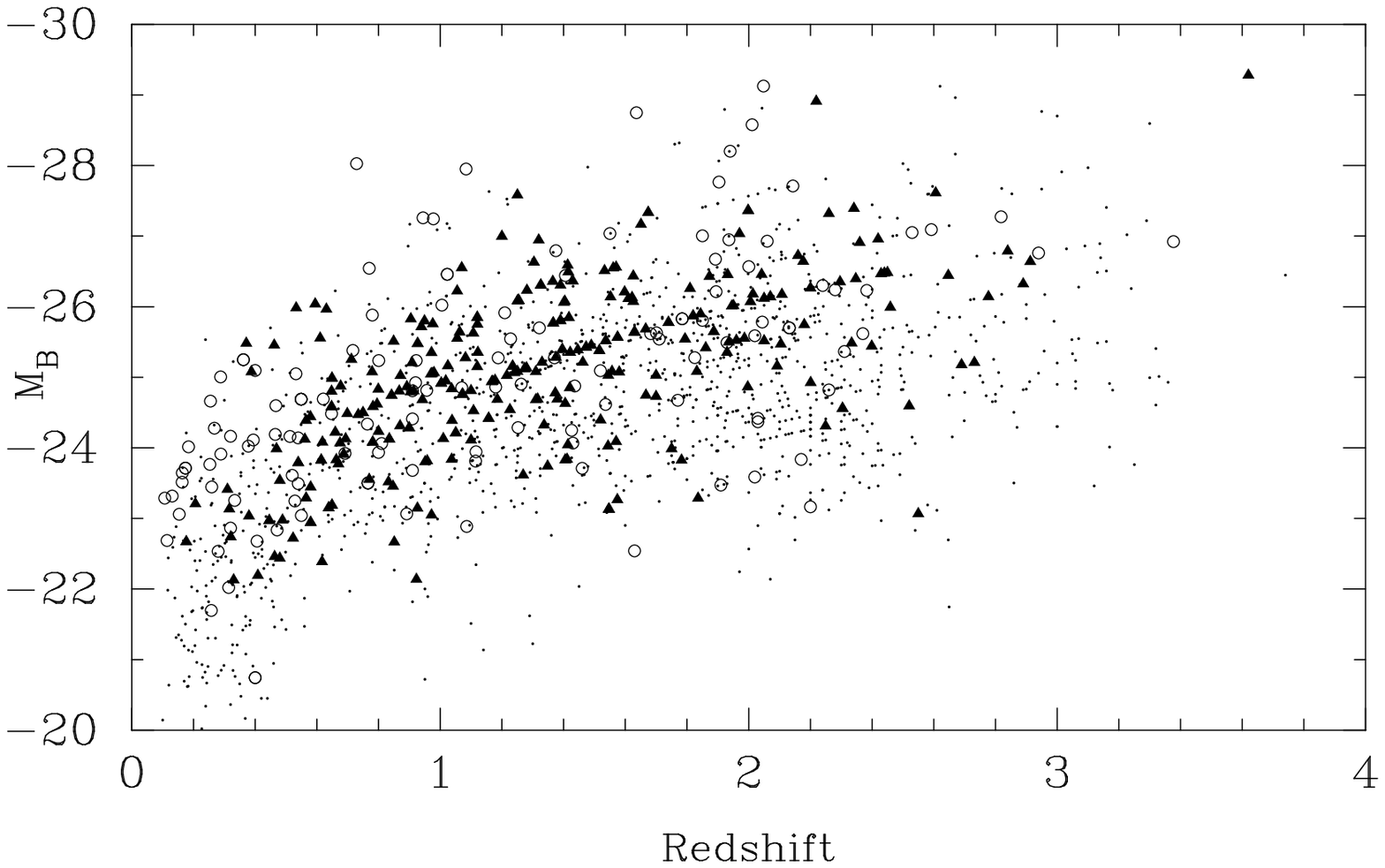}
\caption{Absolute magnitude of quasars in our sample as a function of
redshift. Non-radio selected (mostly optical) quasars with
FIRST detection are indicated by open circles, solid triangles
indicate radio selected quasars. The upper limits are represented by dots.}
\labfig{abmz}
\end{figure}

We have shown in \fig{abmz}  the variation of absolute magnitude with
redshift for all the quasars in our sample.  In this  
figure, triangles represent RSQ while the OSQD are represented by
open circles and the OSQU  by dots.  The  optical luminosity 
of the sample  ranges over $\sim4$ orders of magnitude, and 
the redshift goes upto $\sim3.6$.  All three kinds of quasar
are distributed over much of these wide ranges.   

\subsection{Distribution of radio luminosity}
\labsubsecn{radlum}

The radio-selected quasars (RSQ) in our sample have all been
discovered in radio surveys with flux limits much higher than the
$1\mjy$ limit of the FIRST survey.  For a given redshift,
these quasars will have much higher radio luminosity than most of the
non-radio selected component of our population.  The radio luminosity
distribution of the RSQ is consequently not representative of the
distribution for the overall quasar population.  We shall therefore
omit the RSQ from the following considerations, except where they are
needed in some specific context.

\begin{figure}
\plotone{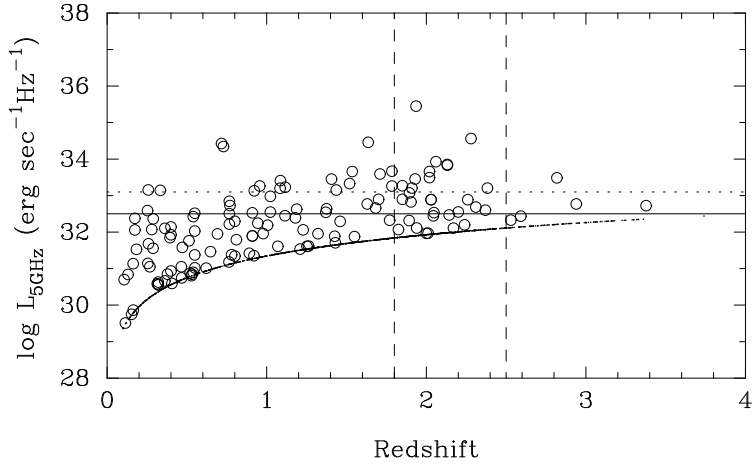}
\caption{Radio luminosity as a function of redshift. The locus of dots 
indicates the 1 mJy upper limits. The horizontal dotted line is the
dividing radio luminosity between radio loud and radio quiet objects
adopted by MPM90. The solid horizontal line is the dividing luminosity
that we have chosen. The region of redshift space explored by MPM90 is 
between the two vertical dashed lines.}
\labfig{lrz}
\end{figure}

\begin{figure}
\plotone{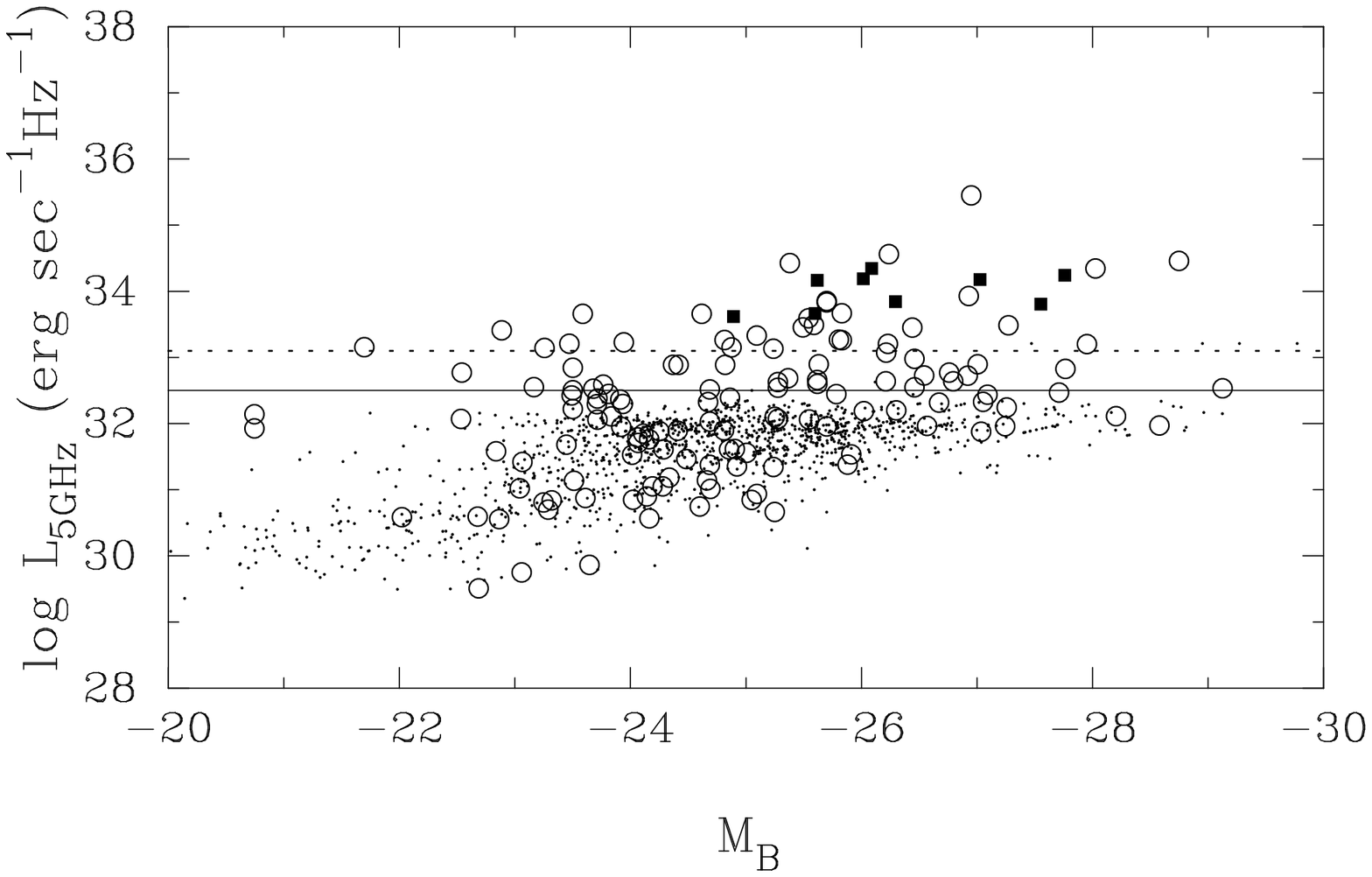}
\caption{Radio luminosity as a function of absolute magnitude. The
filled squares are radio detections of quasars studied in MPM90. The
other symbols are as in \fig{lrz}. The dotted horizontal line is the
MPM90 dividing luminosity between radio loud and radio quiet quasars,
in our units. The solid horizontal line is the dividing luminosity that we have chosen.}
\labfig{lrabm}
\end{figure}

We have shown in \fig{lrz} a plot of the $5\ghz$ radio luminosity
against redshift for non-radio selected quasars.  The OSQD are shown
as unfilled circles, while the OSQU are shown as dots, and form the
almost continuous lower envelope which indicates the radio luminosity
corresponding to a radio flux of $1\mjy$ over the redshift range.  In
\fig{lrabm} is shown a plot of the $5\ghz$ radio luminosity against
the absolute blue magnitude for the non-radio selected quasars.  In
this figure too, radio detections are shown as open circles, and the
radio upper limits as dots.  There appears to be a correlation between
the logarithm of the radio luminosity and absolute magnitude, in spite
of the large scatter in radio luminosity for a given absolute
magnitude. The linear correlation coefficient for the 135 radio
detections alone is $0.22$, which is significant at the $>99.9\percent$
confidence level.  However, it is seen from \fig{abmz} and \fig{lrz} that mean
radio as well as  optical luminosity increase with redshift, which is due to the existence of a limiting radio flux and apparent
magnitude in the surveys in which quasars are discovered. A situation
can arise in which an observed correlation between radio luminosity
and absolute magnitude is mainly due to the dependence of each luminosity 
on the redshift $z$.  It is is important to see if the correlation
remains significant when such an effect of the redshift on
the observed correlation is taken into
account. This can be done by evaluating a {\em partial linear
correlation coefficient} as follows (Havilcek \& Crain 1988; Kembhavi
\& Narlikar 1999). 

Let $r_{L_r,M}$, $r_{L_r,z}$ and
$r_{M,z}$ be the correlation coefficients between the pairs $\log L_r$ 
and $M$, $\log L_r$ and $z$, and $M$ and $z$ respectively. The partial 
linear correlation coefficient is then defined by
\be
r_{L_r,M;z} = \frac{r_{L_r,M}^2 - r_{L_r,z}
r_{M,z}}{\sqrt{1 - r_{L_r,z}^2} \sqrt{1 -  r_{M,z}^2}}
\ee

The partial correlation correlation coefficient has the same
statistical distribution as the ordinary correlation coefficient and
therefore the same tests of significance can be applied to it. A
statistically significant value for it means that the luminosities are
correlated at that level of significance even after accounting for
their individual dependence on the redshift.

For our sample of 135 radio detections, the partial linear correlation
coefficient is $0.09$, which is significant only at the $72\percent$
confidence level.  The observed correlation between the radio
luminosity and absolute magnitude thus appears to be largely induced
by the effect of the large range in redshift over which the sample is
observed.  The lack of correlation found here is consistent with the
results of Miller, Peacock and Mead (1990, hereafter MPM90) and Hooper
\etal (1995).

MPM90 have observed a sample of optically selected quasars, with
redshift in the range $1.8<z<2.5$, with the VLA to a limiting
sensitivity of $\sim 1\mjy$ at $5\ghz$. They detected nine quasars out
of a sample of 44; these objects are shown in \fig{lrabm} as filled
squares.  The radio upper limits of MPM90 occupy the same range as our
upper limits shown in the figure, and are not separately indicated.
MPM90 have commented at length on the luminosity gap found between
their radio detections and upper limits.  They concluded that the gap
was indicative of a bimodality in the distribution of radio
luminosity, which divides quasars into a radio loud population, with
radio luminosity $>\onlyten{25}\whzs$, and a radio quiet population
with luminosity $<\onlyten{24}\whzs$.  The radio loud quasars were
taken to be highly luminous representatives of the population of radio
galaxies, and the radio quiet population was taken to be like Seyfert
galaxies.  The conspicuous gap between radio detections and upper
limits is absent in our data.  It is seen in \fig{lrabm} that the
region $\sim\onlyten{32}\lneq\lr(5\ghz)\lneq\onlyten{33}\ergsh$ (which
corresponds to the gap found by MPM90 for our units and constants) is
occupied by many quasars.  Only seven of these are in the redshift
range of the MPM90 sample, which probably explains why they did not
find any quasars in the gap: our sample is about 30 times larger, and
even then we find only a small number in the range.

\begin{figure}
\epsscale{1.0}
\plotone{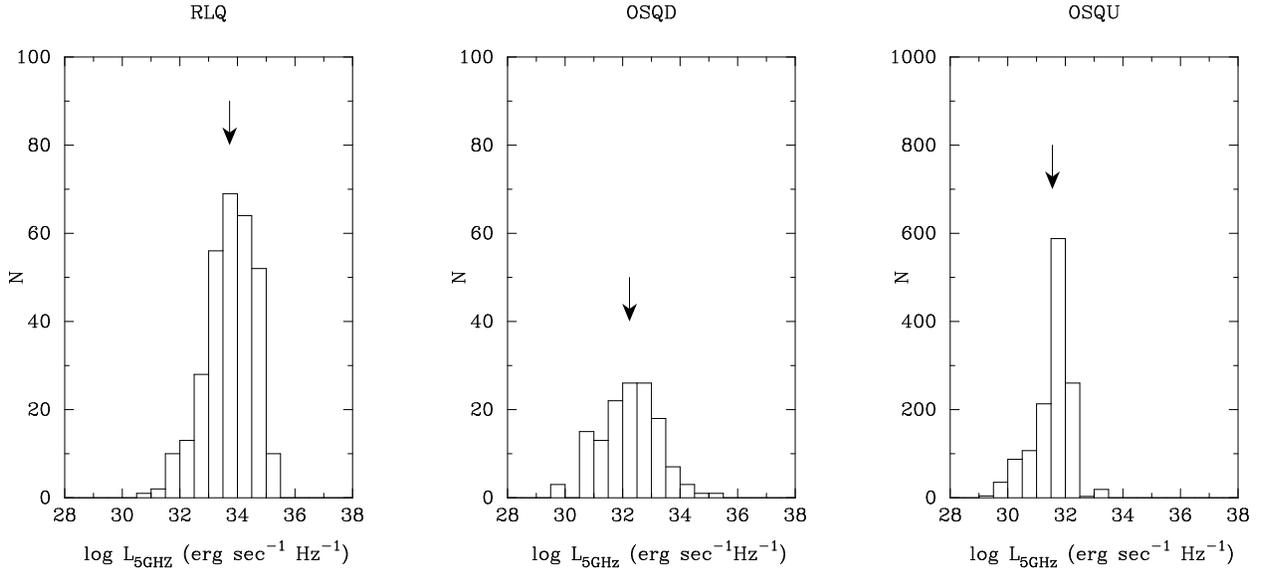}
\caption{Distribution of radio luminosity for the three kinds of
quasars. The arrow indicates the mean value.}
\labfig{lrhist}
\end{figure}

In \fig{lrhist} we show the distribution of the log of  radio
luminosity for the
RSQ, the OSQD and the OSQU. The mean value for each is indicated by
an arrow.
The radio luminosity of the OSQD has a mean value of $\onlyten{32.24}\ergsh$), which is approximately 1.5 orders of
magnitude fainter than the mean luminosity of the RSQ, because the
latter were selected in high flux limit surveys. The RSQ have a median
radio flux of $\sim400\mjy$, while there are only three OSQD with
radio flux $\geq100\mjy$.  The radio luminosity upper limits of the
OSQU are well mixed with the fainter half of the luminosity
distribution of the OSQD.  The rather sharp cutoff in the luminosity
upper limit distribution of the OSQU is due to the flattening in the
$1\mjy$ luminosity envelope in \fig{lrz} at high redshifts.  The upper
limits peak at a luminosity which is approximately half a decade lower
than the peak in the luminosity distribution of the OSQD. The mean
value for the OSQU is $\onlyten{31.55}\ergsh$. A Kolmogorov-Smirnov
test on the distribution of radio luminosity of the OSQD and OSQU 
shows that they are drawn from different distributions with a
significance of $99.9\percent$. This is consistent with a bimodal 
distribution amongst the radio detections and upper
limits. If the radio luminosity distribution is indeed bimodal, the present radio upper limits, when observed to a limiting
flux significantly less than $1\mjy$, would be found to have radio
luminosities considerably lesser than the present set of detections. 

\subsection{Distribution of radio-to-optical luminosity ratio $R$}
\labsubsecn{disr}

The ratio $R$ is defined using rest frame monochromatic radio and
optical luminosities at some fiducial rest frame wavelengths.  In the
following we will choose these to be at $5\ghz$ and $2500\unit{\AA}$
in the radio and optical case respectively.  With our choice of
spectral indices $\ar = \aop = 0.5$, $\log R$ is given in terms of
observed flux densities at observed wavelengths at $5\ghz$ and $2500
\unit{\AA}$ by

\be
\log R = \log \fr(5\ghz) - \log \fop(2500\unit{\AA}).
\labequn{r1}
\ee

\fig{rz} shows the variation of $R$ with redshift. There is
considerable overlap for $R\lneq3$ between the radio detections and
upper limits, but there are only detections at the highest values of
R. There is only one upper limit with $R > 3$.  At each redshift, there
is a maximum to the $R$ upper limits, and this increases
slowly with redshift, so that an envelope is seen.  For an upper limit
to be found above the envelope, it would be necessary to have quasars
at fainter optical magnitudes than are presently to be found in the HB
catalogue.  In the case of the detections, the maximum value $R_{\rm
max}(z) = L_{\rm r,max}(z)/L_{\rm op,min}(z)$ decreases with redshift.
This occurs because the increase in $L_{\rm r,max}(z)$ with redshift
is slower than the increase in $L_{\rm op, min}(z)$ with redshift, as
can be seen from \fig{abmz} and \fig{lrz}.  Similarly, the minimum value of
$R$ for the detections, $R^{\rm min}(z) = L_{\rm r,min}(z)/L_{\rm
op,max}(z)$, increases with redshift, because $L_{\rm op, max}(z)$
increases slower than $L_{\rm r,min}(z)$.

\begin{figure}
\epsscale{0.75}
\plotone{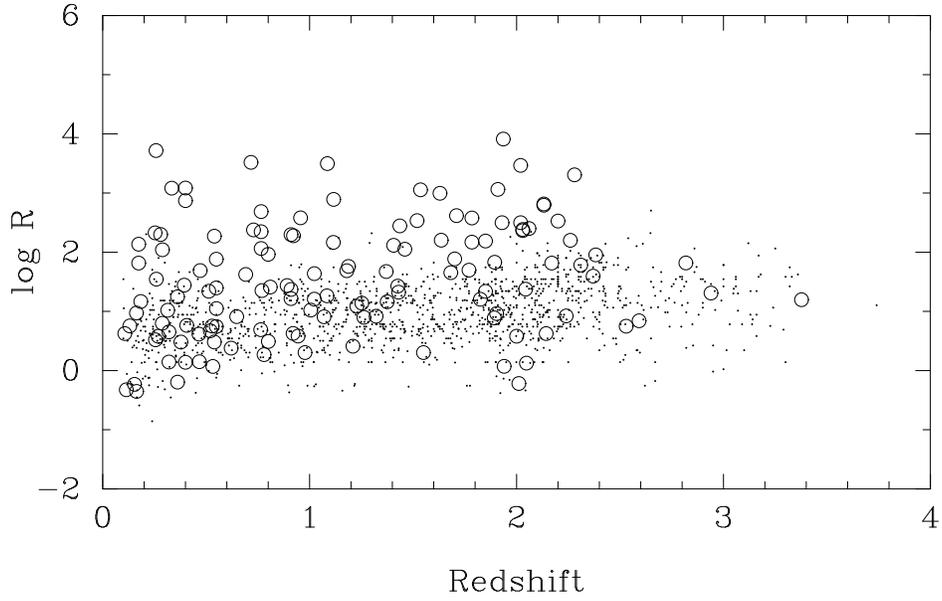}
\caption{$R = \lr/\lop$ as a function of redshift. Symbols are as in \fig{lrz}}
\labfig{rz}
\end{figure}

\begin{figure}
\plotone{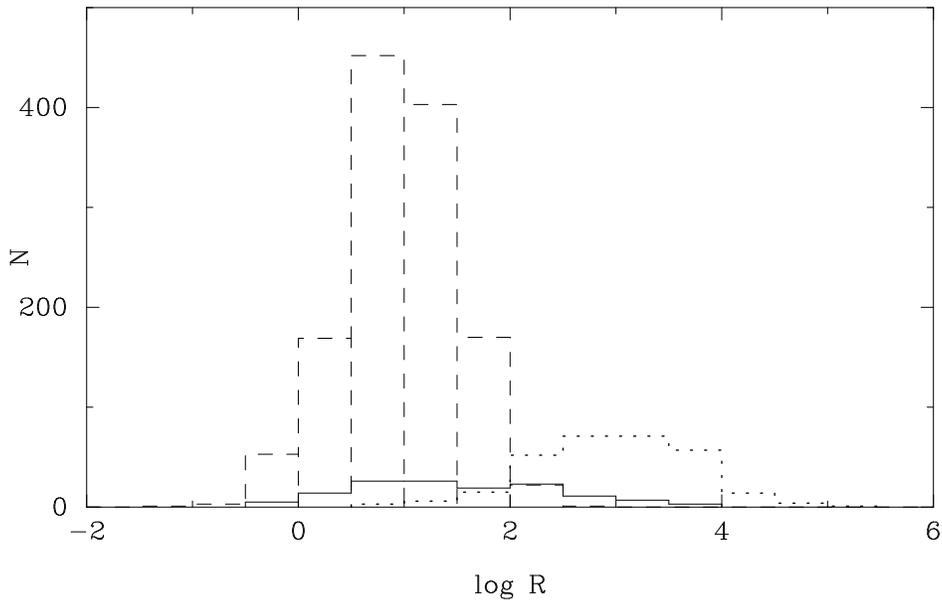}
\caption{Distribution of $R = \lr/\lop$ for quasars with radio
detections (solid line), compared to that for upper limits (dashed
line). Radio selected quasars are shown with a dotted line, for comparison.}
\labfig{rhist}
\end{figure}

\fig{rhist} shows a histogram of $\log R$ for radio detections (solid
line) and radio upper limits (dashed line).  For comparison, the
distribution of $R$ for the radio selected quasars is shown as a
dotted line. An important question here is whether the distribution of
$R$ is bimodal. The number of radio detections is not large enough to
provide information about the distribution of $R$ over its wide range.
However, as mentioned above, there is considerable overlap in the
distributions of the detections and upper limits in the region $0 \le
R \le 3$. It is therefore possible, in principle, to use statistical
techniques from the field of survival analysis (see \eg Feigelson and
Nelson 1985) to determine the underlying distribution for a mixed
sample of detections and upper limits.  If this joint distribution,
and the overall distribution of detections have distinct maxima, then
one could say that the distribution of $R$ amongst all quasars is
bimodal.
  
The appropriate technique to derive the joint distribution would be
the Kaplan-Meier estimator included as part of the ASURV package
(LaValley, Isobe \& Feigelson, 1992). One of the requirements of this
estimator is that the probability that an object is censored (\ie it
has an upper limit), is independent of the value of the censored
variable. If such {\it random censoring} applies to our sample, then
the shape of the observed distribution of $R$ for the detections and
upper limits should be the same, in the region of overlap $0\leq R
\leq 3$. A Kolmogorov-Smirnov test shows that the two distributions
may be considered to be drawn from the same population at only the
$\sim20\percent$ level of significance. Due to the low level of
significance it is not possible to use the Kaplan-Meier estimator, or
another similar to it, to
obtain a joint distribution. A radio survey with a
lower limiting flux than FIRST would be needed to convert the upper
limits to detections and to constrain the distribution of $R$ at its
lower end.  Additional quasars with higher $R$ values can be found by
increasing the area covered by the FIRST survey.

We have mentioned in \subsecn{bivariate} that the separation of the
bivariate luminosity function as in \equn{lumfn2} is most useful if
$R$ is independent of the optical luminosity.
Moreover, such a separation implies that the mean radio luminosity
must increase with the optical luminosity.  Such a correlation between
the luminosities is not seen in \fig{lrabm}, and as discussed in
\subsecn{radlum}, $\lr$ appears to be distributed independently of
$\lop$ (\ie absolute magnitude).  This requires that the distribution
of $R$ depends on $\lop$ and separation as in \equn{lumfn2} is not
possible.  Separation of the bivariate function as in \equn{lumfn1} 
therefore appears to be the preferred alternative.
 
\section{Radio-loud Fraction}

As mentioned in the introduction, the boundary between radio
loud and radio quiet quasars can be defined either (1) in terms of a
characteristic value of the radio to optical luminosity ratio $R$, say
$R=1$, or (2) in terms of a characteristic radio luminosity.  These
two criteria are related to the two ways in which the bivariate
luminosity function can be split up between the optical and radio
parts as discussed in
\subsecn{bivariate}.  We have found no correlation between the radio
and optical luminosities, which implies that a separation involving
$R$, as in \equn{lumfn2} is not consistent with the data. The
distribution of $R$ therefore must be luminosity dependent, and using
a single value of $R$ for separation between radio loud and quiet populations is not
appropriate.  In this situation, we prefer to adopt the criterion for
radio loudness which uses radio luminosity as the discriminant as in
MPM90.

The dividing radio luminosity chosen by MPM90, in our units, is
$\onlyten{33.1}\ergsh$.  This choice was made  on the
basis of a clear separation between radio detections and upper limits
observed by them,
which we do not find, as explained in \subsecn{radlum}.  We have shown
the MPM90 division with a dashed line in \fig{lrabm}.  It is seen that
there is a region below this line with a number of FIRST survey radio detections,
but no upper limits.  It is therefore possible for us to reduce the
dividing luminosity to a level of $\onlyten{32.5}\ergsh$, which is
indicated by a solid line in the figure.  We define as radio loud all
quasars with $\lr(5\ghz)>\onlyten{32.5}\ergsh$, and as radio quiet all
quasars below this limit, even though they may have detectable radio
emission.  The radio loud objects tend to have bright absolute
magnitudes, while a dominating fraction of the radio quite detections
have $M_{\rm B}>-25$.  The faintest of the latter objects could
perhaps be active galaxies like Seyferts, which in the local
neighborhood are known to have lower radio luminosities than radio
galaxies.  The radio loud quasars can be considered to be luminous
counterparts of the radio galaxies, as in the unification model
(Barthel 1989).  If the radio loud and quiet classes indeed represent
such a physical division, then the host galaxies of the former would
perhaps be elliptical, as is the case with radio galaxies, while the
hosts of the quiet objects would be disk galaxies like the Seyferts.
Deep optical and near-IR imaging of different types of quasars would
help in settling this issue.
 
\begin{figure}
\plotone{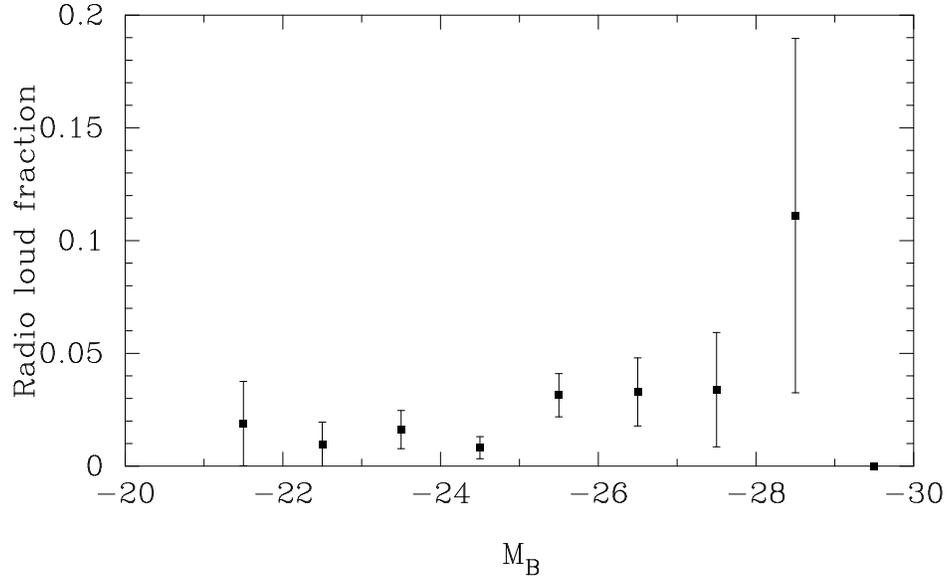}
\caption{Radio loud fraction as a function of absolute
magnitude. The error bar shown is the standard deviation for a random binomial
distribution in the radio detection fraction.}
\labfig{fracabm}
\end{figure}

\begin{figure}
\plotone{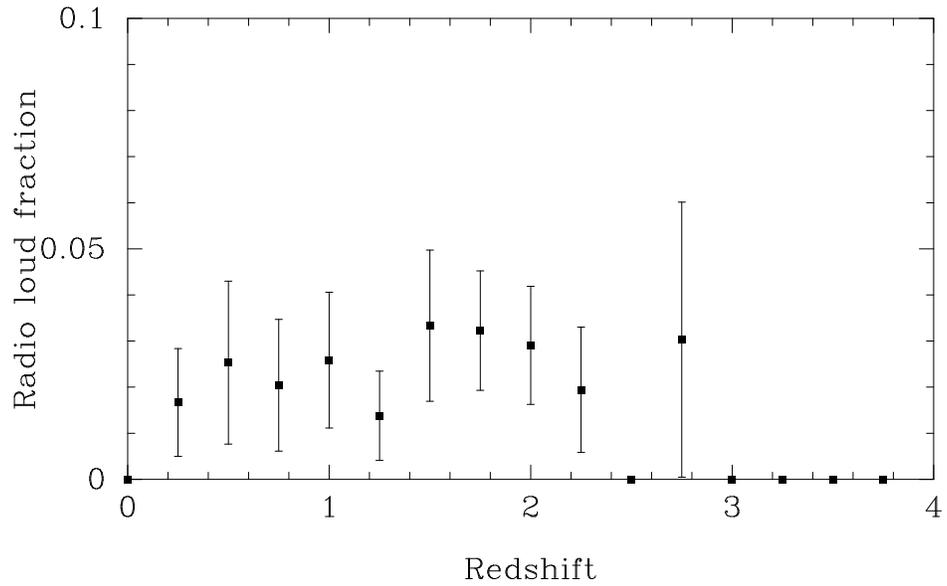}
\caption{Radio loud fraction of quasars as a function of
redshift. The error bars are obtained  as in \fig{fracabm}}
\labfig{fracz}
\end{figure}

We have plotted in \fig{fracabm} the variation of radio loud fraction
of all quasars as a function of absolute magnitude. The fraction here
is taken to be the ratio of the number of radio loud quasars to the
number of all non-radio selected quasars in one absolute magnitude
wide bin.  Each point in \fig{fracabm} is plotted at the centre of the
absolute magnitude bin that it represents.  The error bar shown is the
$\pm1\sigma$ deviation about the detected fraction for a random
binomial distribution in the radio loud fraction.  We find that the
radio loud fraction is independent of the absolute magnitude for
$M_{\rm B}\gneq-25$, while it increases at brighter absolute
magnitudes.  The reason for this is the increase in radio luminosity
towards brighter absolute magnitudes seen in \fig{lrabm}, which arises
due to the existence of optical and radio flux limits and the
consequent redshift dependence of the observed luminosities.  An
explicit dependence of the radio loud fraction on absolute magnitude
would imply a real correlation between the radio and optical
luminosities, which is not consistent with the data as we argued in
\subsecn{radlum}.
 
In \fig{fracz} we have shown the radio loud fraction as a function of
redshift.  Each point in the figure represents quasars in a bin of
width 0.1 in redshift.  The error bars are computed as in
\fig{fracabm}.  In contrast with Hooper \etal (1996), we do not find
a clear peak in the radio loud fraction between a redshift of 0.5
and 1. We find that that the radio loud fraction remains nearly
constant upto a redshift of $z\simeq 2.2$.  There is an indication of
increase in the radio loud fraction at higher redshift, but the number
of objects here is rather small, as is apparent from the size of the
error bars.  A very sharp reduction in the radio loud fraction for
$z<0.5$ was found by BB97. Such a reduction is seen only when radio
selected and non-radio selected quasars are considered together, and
is also apparent in our data if the two kinds of objects are mixed.  
We have chosen not to do that, to keep our results free
from biases introduced by the radio selected objects, as explained in
\subsecn{radlum}.

The large $1\sigma$ error bars on the plots presented in this section,
are caused by the relatively few non-radio selected quasar
detections. Due to these error bars it is not possible to distinguish
unambiguously between alternatives regarding the dependence of radio
loud fraction on other observable properties. More data would be
required to confirm or refute our preliminary conclusions regarding the
evolution of radio loud fraction with absolute magnitude and redshift.

\section{Conclusions}

The main results of our work are:
\begin{itemize}
\item We have reported radio detections of 69 previously undetected
quasars. 
\item We have found additional evidence that the
close pair of quasars 1343+266A and 1343+266B are {\em  not} 
gravitationally lensed.  
\item We have found no correlation between radio luminosity and
optical luminosity for the non-radio selected quasars. Our data is
consistent with a bimodal distribution in radio luminosity.
The distribution of the ratio of radio to optical luminosity  is also
bimodal, but this may have little relevance because of the lack of a
clear correlation between radio and optical luminosities.
\item The  radio loud fraction does not seem to be strongly dependent
on absolute magnitude, which is consistent with the lack of correlation between
radio and optical luminosities.
\item The radio loud fraction does not seem to vary significantly with redshift.
\end{itemize}

The highly heterogeneous nature of the sample used here, makes it
inappropriate for studies in parameter ranges where it is seriously
incomplete, like high redshift radio quasars. Large surveys like the
Digitized Palomar Observatory Sky Survey (DPOSS) and the 
Sloan Digital Sky Survey (SDSS) will remedy this situation, by providing a
large number of quasar candidates for spectroscopic followup.

It is possible that the radio emission from radio loud and radio quiet
quasars may be powered by entirely different physical mechanisms. In
recent years, there have been suggestions that that radio emission in
radio quiet quasars originates in a nuclear starburst rather than
accretion onto a central engine (Terlevich \etal 1992). A logical step
in testing this idea, is to look for differences in the radio
and optical morphology of the quasar environment for the two quasar populations
(eg. Kellerman 1994). We will report work on the radio morphology of
quasar environments obtained from FIRST in a future paper.

\acknowledgements

We thank R. Srianand for helpful comments and discussion. We thank an
anonymous referee whose comments and suggestions helped improve this paper. 

This research
has made use of the NASA/IPAC Extragalactic Database (NED) which is
operated by the Jet Propulsion Laboratory, California Institute of
Technology, under contract with the National Aeronautics and Space
Administration.
\vskip 1cm
\centerline{REFERENCES}
\noindent Barthel, P. D. 1989, \apj, 336, 606\nl
Becker, R. H., Gregg, M. D., Hook, I. M., McMahon, R. G.,
White, R. L. \& Helfand, D. J. 1996, \apjl, 479, L93\nl
Becker, R. H., White, R. L., \& Helfand, D. J. 1995, \apj,
450, 559\nl
Bischof, O. B.,  \& Becker, R. H. 1997 (BB97), \aj, 113, 2000\nl
Brinkmann, W., Yuan, W., Siebert, J. 1997, \aap, 319, 413\nl
Crampton, D. \etal 1988, \apj, 330, 184\nl
Cress, C. M., Helfand, D. J., Becker, R. H., Gregg, M. D., \& White, R. L. 1996, \apj, 473, 7\nl
Condon, J. J., Odell, S. L., Puschell, J. J., Stein, W. A. 1981,
\apj, 246, 624\nl
Condon, J. J., Cotton, W. D., Greisen, E. W., Yin, Q. F., Perley,
R. A., Taylor, G. B. \& Broderick, J. J. 1998, \aj, 115, 1693\nl
Crotts, A. P. S., Bechtold, J, Fang, Y. \& Duncan, R. C. 1994, \apjl,
437, L79\nl
Feigelson, E. D. \& Nelson, P. I. 1985, \apj, 293, 192
Gregg, M. D. Becker, R. H., White, R. L., Helfand, D. J., McMahon,
R. G. \& Hook, I. M. 1996, \aj, 112, 407 \nl
Havilcek, L. L. \& Crain, R. D. 1988, Practical statistics for the
physical sciences (Washington DC: American Chemical Society)\nl
Hewitt A. \& Burbidge G. 1993, \apjs, 87, 451\nl
Hooper, E. J., Impey, C. D., Foltz, C. B. \& Hewett, P. C. 1995,
\apj, 445, 62\nl
Hooper, E. J., Impey, C. D., Foltz, C. B. \& Hewett, P. C. 1996,
\apj, 473, 746\nl
Kellermann, K. I., Sramek, R., Schmidt, M., Shaffer, D. B., \& Green,
R. 1989, \aj, 98, 1195\nl
Kellermann, K. I., Sramek, R. A., Schmidt, M., Green, R. F., \&
Shaffer, D. B. 1994, \aj, 108, 1163\nl
Kembhavi, A. \& Narlikar, J. 1999, Quasars and Active Galactic Nuclei:
an Introduction (Cambridge: Cambridge Univ. Press)\nl
Kukula, M. J., Dunlop, J. S., Hughes, D. H. \& Rawlings, S. 1998,
\mnras, 297, 366\nl
LaValley, M., Isobe, T. \& Feigelson, E. 1992, A.S.P. Conference
Series, Vol. 25,  245\nl
Marshall, H.L. 1987, \apj, 316, 84\nl
Miller, L., Peacock, J. A., \& Mead, A.R.G. 1990, \mnras, 244, 207\nl
Monet, D., Bird, A., Canzian, B., Harris, H., Reid, N., Rhodes, A.,
Sell, S., Ables, H., Dahn, C., Guetter, H., Henden, A., Leggett, S.,
Levison, H., Luginbuhl, C., Martini, J., Monet, A., Pier, J., Riepe,
B., Stone, R., Vrba, F., Walker, R. 1996, USNO-A2.0, (Washington DC: U.S. Naval Observatory)\nl 
Peacock, J. A., Miller, L. \& Longair, M. S. 1986, \mnras, 218, 265\nl
Sramek, R.A. \& Weedman, D.W. 1980,  \apj, 238, 435\nl
Sandage A. 1965,  \apj, 141, 1560\nl
Stocke, J. T., Morris, S. L., Weymann, R. J. \& Foltz,
C. B. 1992, \apj, 396, 487\nl
Terlevich, R., Tenorio-Tagle, G., Franco, J. \& Melnick,
J. 1992, \mnras, 255, 713\nl
Veron-Cetty, M.P., \& Veron, P. 1991, ESO Scientific Report 10\nl
Visnovsky, K. L. \etal 1992,  \apj, 391, 560\nl 
White, R. L., Becker, R. H., Helfand, D. J., \& Gregg, M. D. 1997, \apj, 475, 479\nl

\begin{deluxetable}{l l}
\tablewidth{0pt}
\tablecaption{Summary of radio detections}
\tablehead{\colhead{}& \colhead{}}
\startdata
Number of HB93 quasars in FIRST area:& 1704\nl
Number of quasars with radio detections:& 389\nl
Number of radio selected quasars:& 263\nl
Number of non-radio selected quasars:& 126\nl
Number of non-detections:&1315\nl
Percentage of quasars with detected radio emission:&$\sim$22\%\nl 
Percentage of non-radio selected quasars with detected radio emission:&$\sim$7\%\nl 
\enddata
\end{deluxetable}
\newpage

\begin{deluxetable}{l c c r c l l l}
\tablewidth{0pt}
\tablecaption{FIRST detections of quasars}
\tablehead{\colhead{IAU}& \colhead{Selection \tablenotemark{a}}
& \colhead{$m_{pg}$} & \colhead{1.4 GHz Peak Radio}& \colhead{$z$}
& \colhead{Separation}&\colhead{Alternative}&\colhead{Recent radio}\\
\colhead{Designation}&\colhead{Technique}&&\colhead{Flux(mJy)}&&\colhead{(arcsec)}&\colhead{designation}&\colhead{detection}}

\startdata
0002-018& O&    18.7&  62.26&   1.71& 1.2&&\nl
0003-003& R&    19.35& 3111.27& 1.03& 0.6& 3CR   2&\nl
0004+006& O&    17.8&  1.55&    0.32& 1.3&&\nl
0009-018& O&    18.4&  1.61&    1.07& 2.2& UM    212&\nl
0012-002& O&    17.&   1.45&    1.55& 5.0& UM    221&\nl
0012-004& O&    18.6&  12.67&   1.70& 0.7&&\nl
0013-005& R&    20.8&  1050.26& 1.57& 1.6& PKS&\nl
0019+003& O&    18.6&  1.72&    0.31& 0.5& A&\nl
0020-020& O&    18.4&  8.31&    0.69& 0.4&&\nl
0021-010& O&    18.2&  1.17&    0.76& 0.7&&\nl
0024+003& O&    18.0&  3.50&    1.22& 1.7&&\nl
0029-018& O&    18.7&  13.54&   2.38& 0.3&&\nl
0038-020& RX&   18.5&  593.72&  1.17& 8.0& PKS&\nl
0038-019& RX&   16.86& 272.43&  1.67& 8.4& PKS&\nl
0038-019& RX&   16.86& 441.62&  1.67& 8.4& PKS&\nl
0040+005& O&    18.&   1.09&    2.00& 2.0& UM    269&\nl
0043+008& OXR&  17.&   3.04&    2.14& 1.0& UM    275&\nl
0045-013& O&    18.&   1.61&    2.53& 1.3& UM    278&\nl
0045-000& CR&   19.4&  89.40&   1.53& 1.4& PKS&\nl
0048+004& O&    18.2&  13.70&   1.18& 1.1&&\nl
0052-002& O&    17.7&  3.08&    0.64& 1.1&&\nl
0054-006& R&    19.1&  114.74&  2.77& 0.2& PKS&\nl
0056-001& CXR&  17.02& 2324.98& 0.71& 0.5& PHL   923&\nl
0059-021& O&    18.0&  2.37&    1.32& 0.7&&\nl
0100+004& O&    19.0&  31.65&   1.43& 4.7&&\nl
0101-025& R&    19.1&  259.45&  2.05& 3.6& PKS&\nl
0103-021& R&    19.84& 613.74&  2.20& 1.4& PKS&\nl
0105-008& R&    17.5&  883.60&  0.31& 1.0& PKS&\nl
0107-025& C&    18.2&  90.65&   0.95& 2.3& QSO   10&\nl
0112-017& RX&   17.41& 1025.07& 1.36& 0.3& PKS&\nl
0122-005& O&    18.6&  334.34&  2.28& 1.4& UM    320&\nl
0122-003& R&    16.70& 1481.35& 1.07& 0.5& PKS&\nl
0131+009& OR&   18.5&  8.53&    1.37& 2.9& UM    338&\nl
0133+004& C&    18.5&  4.25&    0.91& 7.1& NGC   622&\nl
0133+004& C&    20.2&  4.25&    1.46& 7.1& NGC   622&\nl
0137-018& O&    18.5&  1.51&    2.24& 1.9& UM    356&\nl
0150-017& O&    19.0&  35.97&   2.02& 1.2& UM    375&BB97\nl
0157+001& CX&   15.69& 22.55&   0.16& 7.9& MKN   1014&BB97\nl
0157+011& R&    18.5&  537.75&  1.17& 1.5& 4C    01.05&\nl
0222+000& R&    19. &  274.34&  0.52& 1.7& PKS&\nl
0222-008& R&    18.4&  220.60&  0.68& 5.9& PKS&\nl
0222-008& R&    18.4&  813.29&  0.68& 6.5& PKS&\nl
0225-014& R&    18.15& 149.85&  2.04& 2.4& PKS&\nl
0236-015& R&    18.80& 76.00&   1.79& 6.9& PKS&\nl
0236-015& R&    18.80& 115.33&  1.79& 7.7& PKS&\nl
0240-021& R&    19.69& 184.75&  0.61& 0.7& PKS&\nl
0241+011& R&    20.  &  39.25&  1.41& 2.4& NGC   1073&\nl
0242+009& R&    19.60& 6.06&    1.52& 0.5& PKS&\nl
0244-019& C&    18.5&  26.78&   1.78& 7.9& US    3148&\nl
0244-019& C&    18.5&  68.33&   1.78& 1.3& US    3148&\nl
0244-012& C&    16.88& 1.13&    0.46& 9.5& US    3150&\nl
0248-001& C&    19.04& 53.39&   0.76& 8.8& US    3224&\nl
0248-001& C&    19.04& 12.66&   0.76& 1.3& US    3224&\nl
0248-001& C&    19.04& 24.34&   0.76& 7.9& US    3224&\nl
0249+007& C&    18.66& 7.63&    0.47& 1.4&&\nl
0251-000& C&    18.59& 7.50&    1.68& 0.5& US    3293&\nl
0252-002& C&    19.61& 1.74&    1.42& 1.2&&\nl
0254+007& C&    19.53& 10.23&   1.11& 7.4&&\nl
0256-021& O&    18.5&  1.05&    0.40& 1.0&&Hooper \etal 1995\nl
0256-000& O&    18.72& 2.32&    3.37& 3.4&&\nl
0256-005& R&    17.20& 5.02&    1.99& 7.4& PKS&\nl
0256-005& R&    17.20& 225.85&  1.99& 5.0& PKS&\nl
0257+004& C&    16.71& 1.10&    0.53& 1.4& US    3472&\nl
0259+010& C&    19.64& 3.15&    1.77& 2.3&&\nl
0300-004& R&    18.2&  622.14&  0.69& 1.7& PKS&\nl
0317-023& R&    19.5&  352.46&  2.09& 6.3& 4C    02.15&\nl
0704+384& R&    17.5&  66.80&   0.57& 0.3& 4C    38.20&\nl
0711+356& R&    18.06& 1506.51& 1.62& 0.9& OI    318&\nl
0714+457& R&    \nodata&  383.82&0.94&9.2& S4&\nl
0726+431& R&    18.5&  100.10&  1.07& 9.0& 4C    43.14&\nl
0726+431& R&    18.5&  160.29&  1.07& 8.2& 4C    43.14&\nl
0729+391& R&    18.4&  117.16&  0.66& 0.5& B3&\nl
0730+257& R&    20.&   338.28&  2.69& 3.1& 4C    25.21&\nl
0731+479& R&    18.&   356.64&  0.78& 1.2& S4&\nl
0738+313& RX&   16.16& 2051.49& 0.63& 0.3& OI    363&\nl
0739+398& R&    19.2&  375.97&  1.70& 1.6& B3&\nl
0740+235& R&    19.&   107.82&  0.77& 2.7& OI    267&\nl
0740+380& RX&   17.6&  1113.79& 1.06& 0.6& 3CR   186&\nl
0742+318& R&    16.&   614.67&  0.46& 0.4& 4C    31.30&\nl
0745+241& R&    19.&   694.04&  0.40& 0.4& B2&\nl
0746+483& R&    18.5&  678.22&  1.95& 0.6& OI    478&\nl
0748+333& R&    18.04& 549.70&  1.93& 0.8& OI    380&\nl
0749+379& R&    16.5&  11.76&   1.20& 3.9& UT&\nl
0750+339& R&    18.5&  43.43&   2.07& 9.4& UT&\nl
0750+339& R&    18.5&  17.31&   2.07& 0.8& UT&\nl
0751+563& O&    19.91& 1.23&    4.28& 1.4& PC&\nl
0751+298& R&    18.5&  398.19&  2.10& 0.6& 4C    29.27&\nl
0752+258& R&    18.41& 50.12&   0.44& 5.4& OI    287&\nl
0752+258& R&    18.41& 234.88&  0.44& 4.8& OI    287&\nl
0759+341& R&    18.5&  43.95&   2.44& 3.9& UT&\nl
0801+303& R&    18.5&  1031.39& 1.44& 0.2& 4C    30.13&\nl
0804+499& R&    17.5&  901.70&  1.43& 1.1& OJ    508&\nl
0805+410& R&    19.&   589.58&  1.42& 1.7& UT&\nl
0808+289& R&    18.8&  39.34&   1.88& 0.2& B2&\nl
0808+289& R&    18.8&  15.32&   1.88& 8.8& B2&\nl
0809+483& RX&   17.79& 7747.95& 0.87& 1.0& 3CR   196&\nl
0810+327& R&    18.&   126.85&  0.84& 0.3& B2&\nl
0810+327& R&    18.&   62.78&   0.84& 7.5& B2&\nl
0812+367& R&    18.&   645.65&  1.02& 2.3& OJ    320&\nl
0812+332& R&    18.&   327.89&  2.42& 0.0& B2&\nl
0814+350& R&    20.0&  6.28&    1.34& 0.1&&\nl
0814+227& R&    18.&   42.17&   0.98& 2.5& 4C    22.20&\nl
0814+425& U&    18.5&  944.26&  0.25& 0.2& OJ    425&\nl
0820+225& R&    19.2&  1919.32& 0.95& 0.7& PKS&\nl
0820+560& R&    18.0&  1363.46& 1.41& 0.3& OJ    535&\nl
0821+394& R&    18.5&  1403.96& 1.21& 0.5& 4C    39.23&\nl
0821+447& R&    18.1&  430.56&  0.90& 3.9& 4C    44.17&\nl
0822+272& C&    17.7&  94.61&   2.06& 1.8& W1&\nl
0824+355& R&    20.5&  913.49&  2.24& 0.6& 4C    35.20&\nl
0827+243& RX&   17.26& 835.48&  0.93& 0.3& OJ    248&\nl
0827+378& R&    18.11& 2032.16& 0.91& 0.1& 4C    37.24&\nl
0829+337& R&    18.5&  211.57&  1.10& 7.0& B2&\nl
0831+349& R&    19.2&  18.30&   1.40& 0.2&&\nl
0832+251& C&    \nodata&  1.84&    0.32&7.3& PG&\nl
0833+276& R&    \nodata&  300.48&  0.76& 6.3& OJ    256&\nl
0833+446& C&    15.51& 9.39&    0.25& 1.5& US    1329&BB97\nl
0834+250& R&    18.&   422.63&  1.12& 0.1& OJ    259&\nl
0838+456& C&    17.39& 65.53&   1.40& 3.7& US    1498&BB97\nl
0841+495& C&    19.&   74.55&   2.13& 1.4& NGC   2639&\nl
0841+495& C&    19.&   70.89&   2.13& 6.2& NGC   2639&\nl
0841+449& O&    20.9&  1.30&    2.17& 9.5&&\nl
0843+349& RX&   18.5&  39.54&   1.57& 0.1&&\nl
0843+349& RX&   18.5&  16.74&   1.57& 7.3&&\nl
0844+446& R&    \nodata&  6.29&    0.46& 6.7& 55W   179&\nl
0849+336& C&    17.4&  1.19&    0.62& 9.2& NGC   2683&\nl
0849+336& C&    18.7&  1.19&    1.26& 9.2& NGC   2683&\nl
0849+336& C&    19.3&  1.19&    1.25& 9.2& NGC   2683&\nl
0850+284& X&    17.7&  71.73&   0.92& 5.1& 1E&\nl
0853+515& C&    19.5&  4.36&    2.31& 7.1& NGC   2693&BB97\nl
0859+470& R&    18.7&  1655.19& 1.46& 1.6& 4C    47.29&\nl
0901+285& R&    17.6&  34.28&   1.12& 0.3& B2&\nl
0904+386& R&    18.5&  39.73&   1.74& 5.2& UT&\nl
0904+386& R&    18.5&  44.26&   1.74& 2.2& UT&\nl
0904+386& R&    18.5&  24.38&   1.74& 9.9& UT&\nl
0906+430& RX&   18.48& 3444.11& 0.67& 0.2& 3CR   216&\nl
0907+381& R&    18.&   250.93&   2.16& 1.7& UT&\nl
0910+392& R&    19.0&  23.71&   0.63& 1.2& B3&\nl
0910+392& R&    19.0&  7.13&    0.63& 4.1& B3&\nl
0913+391& R&    18.5&  967.97&  1.25& 0.0& B3&\nl
0913+391& R&    20.&   967.97&  1.26& 1.2& 4C    38.28&\nl
0917+449& R&    19.&   1079.25& 2.18& 0.5& S4&\nl
0918+381& R&    18.8&  42.90&   1.10& 3.3& B3&\nl
0920+313& R&    18.&   251.68&  0.89& 1.4& B2&\nl
0923+392& RX&   17.86& 2752.50& 0.69& 0.3& 4C    39.25&\nl
0924+301& U&    21.&   52.82&   2.02& 1.1&&\nl
0926+388& R&    18.5&  138.83&  1.63& 0.7& B3&\nl
0927+362& R&    19.&   975.97&  1.15& 4.1& 3CR   220.2&\nl
0927+362& R&    19.&   749.85&   1.15& 3.5& 3CR   220.2&\nl
0928+312& R&    18.6&  126.86&  1.31& 2.9& B2&\nl
0928+349& R&    19.8&  38.65&   0.92& 0.8&&\nl
0928+348& R&    20.3&  11.22&   2.30& 0.4&&\nl
0928+348& R&    20.3&  2.66&    2.30& 8.3&&\nl
0932+367& R&    18.5&  283.00&  2.84& 1.7& UT&\nl
0935+430& C&    18.83& 3.16&    2.04& 1.0& US    795&BB97\nl
0937+391& R&    18.&   41.28&    0.61&6.4&4C    39.27&\nl
0938+450& C&    18.7&  13.81&   0.80& 1.1& US    844&BB97\nl
0941+522& R&    18.6&  628.01&  0.56& 2.1& OK    568&\nl
0941+261& R&    18.7&  730.69&  2.91& 0.6& OK    270&\nl
0945+436& C&    17.78& 2.72&    1.89& 0.3& US    987&BB97\nl
0945+408& R&    17.5&  1439.97& 1.25& 0.7& 4C    40.24&\nl
0949+363& R&    18.5&  99.80&   2.05& 1.7& UT&\nl
0952+441& C&    17.28& 2.30&    0.46& 1.0& US    1101&\nl
0952+457& C&    16.76& 31.38&   0.25& 2.7& US    1107&Brinkmann \etal 1997\nl
0952+357& R&    18.5&  190.74&  1.24& 4.7& 4C    35.21&\nl
0953+254& RX&   17.13& 1041.77& 0.71& 0.2& OK    290&\nl
0954+556& R&    17.7&  2804.17& 0.90& 5.0& PKS&\nl
0955+387& R&    20.0&  161.59&  1.40& 0.2& B3&\nl
0955+476& R&    18.&   763.01&  1.87& 3.9& OK    492&\nl
0955+326& RX&   15.78& 1204.19& 0.53& 0.6& TON   469&\nl
0957+561& R&    17.25& 283.96&  1.41& 3.7& A&\nl
0957+561& R&    17.35& 283.96&  1.41& 7.4& B&\nl
1001+226& R&    18.&   33.93&   0.97& 0.9& 4C    22.26&\nl
1007+417& R&    16.5&  258.74&  0.61& 1.1& 4C    41.21&\nl
1009+334& R&    17.5&  172.71&  2.26& 2.0& UT&\nl
1010+350& R&    19.8&  348.65&  1.41& 0.4& B2&\nl
1011+250& CXR&  15.4&  500.27&  1.63& 1.0& TON   490&\nl
1011+280& R&    18.6&  82.99&   0.89& 4.1& 4C    28.25&\nl
1011+280& R&    18.6&  241.63&  0.89& 5.2& 4C    28.25&\nl
1012+232& R&    17.5&  673.56&  0.56& 0.5& 4C    23.24&\nl
1015+277& R&    17.5&  844.36&  0.46& 8.7& B2&\nl
1015+359& R&    19.&   571.31&  1.22& 4.8& OL    326&\nl
1015+383& R&    18.&   5.39&    0.38& 7.0& UT&\nl
1018+348& R&    17.75& 317.10&  1.40& 0.6& OL    331&\nl
1019+309& R&    16.75& 907.01&  1.31& 0.5& OL    333&\nl
1020+400& R&    17.5&  807.88&  1.25& 1.5& UT&\nl
1028+313& RX&   16.71& 57.04&   0.17& 6.4& B2&\nl
1028+313& RX&   16.71& 58.74&   0.17& 0.3& B2&\nl
1030+415& R&    18.2&  406.68&  1.12& 3.1& VR10.&\nl
1038+528& R&    17.4&  414.78&  0.67& 0.3& OL    564&\nl
1038+528& R&    18.5&  101.76&  2.29& 0.1& B&\nl
1042+349& R&    18.5&  40.45&   2.34& 0.3&&\nl
1044+476& R&    18.4&  734.01&  0.80& 1.4& OL    474&\nl
1045+350& R&    20.8&  17.23&   0.92& 0.3&&\nl
1048+347& R&    20.45& 540.40&  2.52& 2.9& B2&\nl
1048+240& R&    18.5&  282.73&  1.27& 7.9& 4C    24.23&\nl
1048+240& R&    18.5&  110.43&  1.27& 7.0& 4C    24.23&\nl
1050+542& R&    18.2&  117.86&  1.00& 6.5&&\nl
1050+542& R&    18.2&  51.44&   1.00& 5.4&&\nl
1055+499& R&    19.5&  225.14&  2.39& 0.2& 5C2.5&\nl
1059+282& R&    19.&   240.69&  1.86& 0.6& GC&\nl
1105+392& R&    18.5&  603.57&  0.78& 0.7& B3&\nl
1105+392& R&    18.5&  22.78&   0.78& 7.4& B3&\nl
1109+357& X&    18.1&  4.29&    0.91& 6.1& 1E&\nl
1109+350& R&    18.5&  181.57&  1.94& 0.1& UT&\nl
1111+408& RX&   17.98& 1740.12& 0.73& 1.6& 3CR   254&\nl
1115+536& R&    18.4&  612.32&  1.23& 3.0& OM    525&\nl
1115+536& R&    18.4&  266.32&  1.23& 5.7& OM    525&\nl
1115+407& CX&   16.02& 1.04&    0.15& 5.5& PG&\nl
1123+441& R&    19.1&  87.92&   0.48& 0.9& W1&\nl
1123+264& R&    17.5&  904.38&  2.34& 0.1& PKS&\nl
1123+434& R&    18.4&  24.75&   2.01& 2.0& W1&\nl
1124+571& R&    19.0&  473.93&  2.89& 2.4& OM    540/4&\nl
1124+271& C&    17.0&  2.18&    0.37& 2.9& US    2450&BB97\nl
1128+315& C&    16.53& 121.84&  0.28& 3.2& B2&\nl
1130+284& C&    17.52& 9.73&    0.51& 1.9& US    2599&BB97\nl
1132+303& R&    18.24& 306.09&  0.61& 0.6& 3C    261&\nl
1132+303& R&    18.24& 286.52&  0.61& 8.0& 3C    261&\nl
1134+349& R&    19.2&  23.41&   0.83& 0.6&&\nl
1145+321& C&    17.14& 15.76&   0.54& 9.5& US    2978&BB97\nl
1145+321& C&    17.14& 48.24&   0.54& 0.6& US    2978&BB97\nl
1145+321& C&    17.14& 3.51&    0.54& 9.7& US    2978&BB97\nl
1146+562& R&    19.2&  21.88&   0.95& 4.3& W1&\nl
1147+339& R&    18.5&  98.64&   1.49& 0.4& UT&\nl
1148+568& R&    20.5&  83.16&   1.78& 0.7& W1&\nl
1148+477& R&    18.0&  143.90&  0.86& 2.2& 4C    47.33&\nl
1148+549& CR&   15.82& 4.30&    0.97& 0.8& PG&\nl
1148+387& R&    17.04& 391.60&  1.30& 1.2& 4C    38.31&\nl
1150+497& CR&   17.50& 548.11&  0.33& 0.0& LB    2136&\nl
1153+534& R&    20.3&  8.03&    1.75& 1.9& W1&\nl
1153+317& R&    18.96& 2833.64& 1.55& 0.3& 4C    31.38&\nl
1156+295& CR&   14.41& 1855.80& 0.72& 0.1& 4C    29.45&\nl
1157+532& R&    19.7&  129.83&  1.99& 0.7& W2&\nl
1204+399& R&    18.5&  235.09&  1.33& 2.2& UT&\nl
1204+281& R&    18.1&  596.14&  2.17& 2.1& B2&\nl
1206+439& R&    18.42& 1439.20& 1.39& 4.6& 3CR   268.4&\nl
1206+439& R&    18.42& 419.00&  1.39& 4.8& 3CR   268.4&\nl
1207+398& R&    19.4&  23.04&   2.33& 1.2& W3&\nl
1208+322& R&    16.&   19.15&   0.38& 2.5& B2&\nl
1211+334& R&    17.89& 1372.84& 1.59& 8.4& ON    319&\nl
1213+350& R&    20.1&  1323.86& 0.85& 0.3& 4C    35.28&\nl
1214+348& R&    18.7&  154.45&  2.64& 0.6&&\nl
1214+474& R&    19.2&  94.56&   1.10& 0.0& W2&\nl
1215+333& R&    17.5&  183.25&  2.60& 0.6& GC&\nl
1216+487& R&    18.5&  659.98&  1.07& 1.2& ON    428&\nl
1218+339& R&    18.61& 586.79&  1.51& 4.5& 3CR   270.1&\nl
1218+339& R&    18.61& 1929.95& 1.51& 3.8& 3CR   270.1&\nl
1220+373& R&    18.6&  24.36&   0.48& 0.1& B2&\nl
1222+228& CXR&  15.49& 3.86&    2.04& 0.1& TON   1530&\nl
1223+252& CXR&  16.&   6.87&    0.26& 0.6& TON   616&\nl
1225+317& RX&   15.87& 315.34&  2.21& 1.8& B2&\nl
1229+405& R&    19.0&  46.38&   0.64& 7.7& B3&\nl
1229+405& R&    19.0&  186.72&  0.64& 2.7& B3&\nl
1231+349& R&    19.3&  5.85&    0.84& 0.2&&\nl
1231+294& C&    16.&   1.09&    2.01& 3.9& CSO   151&\nl
1234+335& R&    18.5&  175.82&  1.28& 2.2& UT&\nl
1234+265& O&    21.6&  3.50&    2.20& 3.3&&BB97\nl
1240+381& R&    19.&   536.82&  1.31& 0.7& B2&\nl
1244+324& R&    17.2&  77.07&   0.94& 1.4& 4C    32.41&\nl
1247+450& R&    17.8&  338.17&  0.79& 7.3& 4C    45.26&\nl
1248+350& R&    20.0&  240.73&  0.97& 0.3&&\nl
1250+568& RX&   17.93& 2258.50& 0.32& 1.2& 3CR   277.1&\nl
1250+313& O&    16.7&  1.76&    0.78& 0.2& LB    11408&\nl
1251+398& R&    19.2&  30.40&   2.10& 0.9& B3&\nl
1254+370& CR&   17.84& 65.93&   0.28& 0.4& B     142&\nl
1256+357& CXR&  18.24& 15.55&   1.89& 0.9& B     194&\nl
1257+346& CR&   16.99& 10.56&   1.37& 0.5& B     201&\nl
1258+287& RX&   17.38& 192.10&  0.64& 2.9& 5C4.1&\nl
1258+404& R&    19.44& 269.63&  1.66& 6.3& 3CR   280.1&\nl
1258+286& RX&   19.&   78.54&   1.37& 4.3& 5C4.1&\nl
1258+342& OR&   19.&   35.99&   1.93& 5.9& KP    33&\nl
1301+295& CR&   18.9&  42.80&   1.51& 5.8& 5C4.1&\nl
1305+364& CR&   18.01& 1.20&    0.92& 0.8& B     330&\nl
1306+274& R&    18.5&  92.39&   1.53& 8.1& OP    211&\nl
1306+274& R&    18.5&  116.88&  1.53& 2.0& OP    211&\nl
1308+284& O&    18.1&  1.23&    0.52& 4.2& US    370&\nl
1308+297& O&    17.4&  10.82&   1.85& 1.1&&BB97\nl
1309+378& CR&   17.65& 1.20&    0.54& 0.6& B     503&\nl
1309+355& CR&   15.45& 43.92&   0.18& 0.2& PG&\nl
1315+346& R&    19.&   420.95&  1.05& 0.4& OP    326&\nl
1315+346& R&    19.&   29.61&   1.05& 4.1& OP    326&\nl
1315+473& O&    18.01& 1.97&    2.59& 3.6& PC&BB97\nl
1316+269& O&    21.0&  20.84&   1.91& 1.5&&BB97\nl
1316+270& O&    20.0&  7.22&    2.26& 3.9&&BB97\nl
1317+380& R&    18.6&  131.81&  0.83& 3.5& B3&\nl
1317+380& R&    18.6&  70.96&   0.83& 6.5& B3&\nl
1317+520& R&    17.&   297.21&  1.05& 2.8& 4C    52.27&\nl
1328+254& RX&   17.67& 6826.39& 1.05& 0.0& 3CR   287&\nl
1328+307& RX&   17.25& 14777.9& 0.84& 0.2& 3CR   286&\nl
1332+552& R&    16.&   9.88&    1.25& 0.9& 4C    55.27&\nl
1333+459& R&    18.5&  262.72&  2.45& 0.2& S4&\nl
1333+277& O&    19.4&  61.56&   1.11& 4.6&&\nl
1334+246& U&    15.&   19.18&   0.10& 0.2&&\nl
1335+283& O&    20.4&  98.85&   1.08& 0.5&&\nl
1336+351& R&    20.0&  104.72&  1.54& 2.5&&\nl
1338+394& R&    19.0&  14.92&   0.58& 5.4& B3&\nl
1338+394& R&    19.0&  18.12&   0.58& 6.5& B3&\nl
1339+287& R&    18.6&  1.54&    0.33& 1.8&&\nl
1340+287& R&    18.35& 65.57&   1.03& 1.5& B2&\nl
1340+289& R&    17.07& 217.37&  0.90& 1.5& B2&\nl
1342+264& O&    18.6&  8.03&    1.18& 5.4&&BB97\nl
1342+389& R&    17.5&  159.53&  1.53& 4.7& B3&\nl
1343+267& O&    19.8&  1.49&    0.89& 4.2&&\nl
1343+266& O&    20.23& 8.90&    2.03& 7.4&A Crotts \etal 1994&\nl
1343+266& O&    20.18& 8.90&    2.03& 2.1&B Crotts \etal 1994&\nl
1343+386& R&    18.5&  845.79&  1.84& 0.6& 4C    38.37&\nl
1344+264& O&    19.1&  1.66&    1.82& 6.9&&\nl
1347+539& R&    17.3&  960.41&  0.97& 1.2& 4C    53.28&\nl
1348+384& R&    18.&   77.34&   1.39& 0.1& UT&\nl
1348+384& R&    18.&   32.95&   1.39& 8.5& UT&\nl
1348+392& R&    19.0&  130.30&  1.58& 0.5& B3&\nl
1351+267& R&    17.18& 22.35&   0.31& 1.1& B2.2&\nl
1351+318& R&    17.4&  74.13&   1.32& 4.0& B2&\nl
1351+318& R&    17.4&  76.61&   1.32& 4.9& B2&\nl
1353+306& R&    18.2&  123.06&  1.01& 0.9& B2&\nl
1354+258& R&    18.5&  173.68&  2.00& 3.7& OP    291&\nl
1402+436& U&    16.5&  1.59&    0.32& 1.2& CSO   409&\nl
1402+261& CXR&  15.57& 1.19&    0.16& 8.1& PG&\nl
1407+265& CXR&  15.73& 8.85&    0.94& 0.7& PG&\nl
1409+344& R&    18.5&  41.68&   1.82& 7.9& UT&\nl
1409+344& R&    18.5&  92.49&   1.82& 2.7& UT&\nl
1413+373& R&    18.&   406.30&  2.36& 3.7& UT&\nl
1414+347& R&    18.&   60.69&   0.75& 8.2& UT&\nl
1414+347& R&    18.&   31.71&   0.75& 3.6& UT&\nl
1415+451& C&    15.74& 1.09&    0.11& 1.7& PG&\nl
1415+463& R&    17.9&  696.65&  1.55& 0.4& 4C    46.29&\nl
1417+385& R&    19.3&  651.90&  1.83& 0.7& UT&\nl
1419+315& R&    20.90& 78.08&   1.54& 8.5& B2&\nl
1419+315& R&    20.90& 81.43&   1.54& 0.7& B2&\nl
1419+315& R&    20.90& 104.24&  1.54& 7.7& B2&\nl
1421+330& C&    16.70& 8.71&    1.90& 0.4& MKN   679&\nl
1421+359& R&    17.5&  71.37&   1.57& 2.4& UT&\nl
1422+231& R&    16.5&  273.42&  3.62& 0.4&&\nl
1423+242& R&    17.2&  394.80&  0.64& 9.6& 4C    24.31&\nl
1423+242& R&    17.2&  121.14&  0.64& 1.2& 4C    24.31&\nl
1425+267& CXR&  15.68& 1.55&    0.36& 8.5& TON   202&\nl
1425+267& CXR&  15.68& 42.72&   0.36& 0.3& TON   202&\nl
1426+295& R&    18.5&  402.07&  1.42& 0.8& B2&\nl
1435+315& R&    18.&   13.94&   1.36& 3.3& B2&\nl
1435+315& R&    18.&   60.37&   1.36& 1.0& B2&\nl
1435+383& R&    18.&   180.04&  1.61& 1.9& UT&\nl
1435+248& RX&   19.&   252.95&  1.01& 1.0& 4C    24.32&\nl
1435+355& R&    18.&   14.97&   0.54& 9.3& UT&\nl
1435+355& R&    18.&   16.53&   0.54& 4.5& UT&\nl
1441+522& R&    19.97& 1008.23& 1.57& 5.3& 3C    303C&\nl
1444+417& R&    18.2&  73.98&   0.67& 3.6& B3&\nl
1452+301& R&    18.5&  650.82&  0.58& 2.2& OQ    287&\nl
1455+348& R&    20.0&  231.69&  2.73& 0.1&&\nl
1506+339& R&    18.5&  130.95&  2.20& 1.2& UT&\nl
1512+370& RX&   15.5&  48.97&   0.37& 0.3& 4C    37.43&\nl
1520+344& R&    19.&   176.03&  1.31& 2.2& UT&\nl
1522+259& C&    18.79& 1.54&    0.55& 0.7& LB    9695&\nl
1525+314& R&    19.1&  792.83&  1.38& 7.2& B2&\nl
1525+227& CXR&  16.39& 267.42&  0.25& 4.6& LB    9743&\nl
1538+477& CR&   16.01& 40.36&   0.77& 0.9& PG&\nl
1541+355& R&    19.5&  120.89&  1.70& 1.3& UT&\nl
1542+373& R&    17.7&  602.99&  0.97& 0.9& 4C    37.45&\nl
1543+489& C&    16.05& 2.38&    0.40& 4.0& PG&\nl
1546+353& R&    18.&   140.94&  0.48& 1.5& UT&\nl
1555+332& RX&   18.3&  77.06&   0.94& 1.0& GC&\nl
1556+335& RX&   17.&   142.96&  1.65& 0.2& GC&\nl
1605+355& R&    18.&   97.91&   0.97& 2.7& UT&\nl
1606+289& RX&   19.&   3.35&    1.98& 3.5& 4C    28.40&\nl
1611+343& RX&   17.76& 3532.04& 1.40& 0.5& DA    406&\nl
1612+378& R&    18.5&  93.29&   1.63& 7.6& UT&\nl
1612+378& R&    18.5&  46.53&   1.63& 6.5& UT&\nl
1612+261& CXR&  15.41& 17.69&   0.13& 0.5& TON   256&\nl
1620+356& R&    18.5&  177.51&  1.47& 9.2& 4C    35.41&\nl
1620+356& R&    18.5&  10.79&   1.47& 2.2& 4C    35.41&\nl
1621+392& R&    17.5&  189.57&  1.97& 1.9& UT&\nl
1621+361& R&    18.5&  259.58&  0.87& 1.4& UT&\nl
1622+238& RX&   17.47& 635.34&  0.92& 7.8& 3CR   336&\nl
1622+238& RX&   17.47& 101.74&  0.92& 4.1& 3CR   336&\nl
1622+395& R&    17.5&  76.71&   1.12& 5.4& UT&\nl
1622+395& R&    17.5&  132.32&  1.12& 3.5& UT&\nl
1623+269& RX&   17.5&  368.12&  0.77& 1.3& 4C    26.48&\nl
1624+416& R&    22.&   1694.61& 2.55& 0.0& 4C    41.32&\nl
1624+349& R&    19.4&  26.35&   1.33& 0.8&&\nl
1628+380& O&    17.0&  20.00&   0.39& 1.1&&\nl
1628+363& R&    17.5&  149.78&  1.25& 5.4& 4C    36.28&\nl
1628+363& R&    17.5&  52.46&   1.25& 1.0& 4C    36.28&\nl
1628+363& R&    17.5&  211.31&  1.25& 9.7& 4C    36.28&\nl
1629+439& R&    18.5&  581.05&  1.16& 0.9& 4C    43.39&\nl
1631+373& O&    18.6&  3.37&    2.94& 1.3&&\nl
1631+395& O&    16.7&  41.29&   1.02& 0.1&&\nl
1631+395& O&    16.7&  15.24&   1.02& 8.8&&\nl
1632+391& R&    18.&   915.40&  1.08& 0.6& 4C    39.46&\nl
1633+382& RX&   18.1&  2653.87& 1.81& 1.8& GC&\nl
1634+269& R&    17.75& 17.29&   0.56& 7.4& PKS&\nl
1636+473& R&    \nodata&  601.81&  0.74&8.8& 4C    47.44&\nl
1638+398& R&    18.5&  1088.22& 1.66& 1.5& NRAO  512&\nl
1640+396& XR&   18.3&  40.63&   0.54& 6.6&&\nl
1640+401& XR&   17.1&  6.89&    1.00& 8.7&&\nl
1641+399& RX&   15.96& 6050.06& 0.59& 0.4& 3CR   345&\nl
1656+348& R&    19.&   406.35&   1.93& 0.3& OS    392&\nl
1656+571& R&    17.4&  817.61&  1.28& 1.5& 4C    57.28&\nl
1656+477& R&    18.0&  873.76&  1.62& 0.1& S4&\nl
1657+265& R&    18.&   391.07&  0.79& 1.3& 4C    26.51&\nl
1700+518& C&    15.43& 19.20&   0.28& 0.9& PG&\nl
1701+379& R&    19.&   83.11&   2.45& 6.5& UT&\nl
1702+298& R&    19.14& 1200.93& 1.93& 0.3& 4C    29.50&\nl
1705+456& R&    17.6&  681.65&  0.64& 0.7& 4C    45.34&\nl
1705+456& R&    17.6&  10.17&   0.64& 8.3& 4C    45.34&\nl
1710+329& R&    19.&   167.28&  1.96& 1.4& UT&\nl
1713+504& R&    \nodata&  44.92&   1.09& 9.5& 53W   009&\nl
1714+502& R&    \nodata&  47.66&   1.12& 9.7& 53W   015&\nl
1715+535& CR&   16.30& 1.62&    1.94& 4.4& PG&\nl
1718+481& CR&   15.33& 61.37&   1.08& 0.6& PG&\nl
1719+357& R&    \nodata&  386.51&  0.26& 3.4& B2&\nl
1719+497& R&    \nodata&  97.24&   2.15& 7.3& 53W   075&\nl
1719+348& R&    21.1&  55.23&   1.83& 0.5&&\nl
1720+499& R&    \nodata&  10.15&   0.54& 8.4& 53W   080&\nl
1720+499& R&    \nodata&  7.25&    0.54& 7.4& 53W   080&\nl
1720+499& R&    \nodata&  5.27&    1.82& 6.2& 53W   085&\nl
1721+343& RX&   16.5&  438.57&  0.20& 0.5& 4C    34.47&\nl
1724+399& R&    18.&   475.49&  0.66& 0.8& UT&\nl
1726+344& R&    18.5&  72.17&   2.42& 0.8& UT&\nl
1727+386& R&    17.5&  240.27&  1.39& 1.8& UT&\nl
1729+491& R&    18.8&  782.21&  1.03& 0.4& 4C    49.29&\nl
1729+501& R&    17.7&  50.10&   1.10& 0.6& 4C    50.43&\nl
1738+499& R&    19.&   409.90&  1.54& 0.3& OT    463&\nl
1739+522& RX&   18.5&  1508.16& 1.37& 1.5& 4C    51.37&\nl
2131-009& XR&   21.6&  10.30&   1.63& 0.8&&\nl
2134+004& CXR&  17.55& 3546.71& 1.93& 0.9& PHL   61&\nl
2211+006& O&    19.23& 18.31&   0.91& 6.3& PC&\nl
2227-088& R&    17.5&  952.78&  1.56& 0.5& PKS&\nl
2231-008& O&    17.6&  1.07&    1.20& 1.4&&\nl
2235+009& O&    18.5&  1.02&    0.52& 1.0&&\nl
2245-009& O&    17.4&  1.54&    0.80& 1.7&&\nl
0041+001& R& 19.28& 108.53& 1.12& 0.0& PKS \nl
0742+333& R& 17.7& 98.52& 0.61& 0.2& GC \nl
0952+338& C& 17.& 35.73& 2.50& 0.1& CSO 239\nl
1255+370& R& 17.8& 690.34& 0.28& 0.8& B2 \nl
1339+274& O& 19.0& 238.13& 1.18& 0.3&  \nl
1343+284& O& 18.0& 5.83& 0.65& 0.2&  \nl
1420+326& R& 17.5& 412.83& 0.68& 0.2& OQ 334\nl
1623+268& O& 17.3& 10.12& 2.52& 0.2& KP 77\nl
\enddata
\tablenotetext{a}{Selection Technique O:Objective Prism R: Radio C: UV-Excess
X: X-Ray U: Selection technique not mentioned.}
\end{deluxetable}

\end{document}